\documentclass[iop]{emulateapj}

\shorttitle{\emph{NuSTAR} Observations of NGC 5643.}
\shortauthors{A. Annuar et al.}

\usepackage{color}

\begin{document}

\title{\emph{N\MakeLowercase{u}STAR} Observations of the Compton-thick Active Galactic Nucleus and Ultraluminous X-ray Source Candidate in NGC 5643}

\author{A. Annuar\altaffilmark{1}, P. Gandhi\altaffilmark{1,2}, D. M. Alexander\altaffilmark{1}, G. B. Lansbury\altaffilmark{1}, P. Ar\'{e}valo\altaffilmark{3,4}, D. R. Ballantyne\altaffilmark{5}, M. Balokovi\'{c}\altaffilmark{6}, F. E. Bauer\altaffilmark{4,7,8,9}, S. E. Boggs\altaffilmark{10}, W. N. Brandt\altaffilmark{11,12,13}, M. Brightman\altaffilmark{6}, F. E. Christensen\altaffilmark{14}, W. W. Craig\altaffilmark{10,15}, A. Del Moro\altaffilmark{1}, C. J. Hailey\altaffilmark{16}, F. A. Harrison\altaffilmark{6}, R. C. Hickox\altaffilmark{17}, G. Matt\altaffilmark{18}, S. Puccetti\altaffilmark{19,20}, C. Ricci\altaffilmark{21}, J. R. Rigby\altaffilmark{22}, D. Stern\altaffilmark{23}, D. J. Walton\altaffilmark{6,24}, L. Zappacosta\altaffilmark{20}, and W. Zhang\altaffilmark{22}.}

\affil{$^{1}$Centre for Extragalactic Astronomy, Department of Physics, University of Durham, South Road, Durham, DH1 3LE, UK}
\affil{$^{2}$Department of Physics $\&$ Astronomy, Faculty of Physical Sciences and Engineering, University of Southampton, Southampton, SO17 1BJ, UK}
\affil{$^{3}$Instituto de F\'{i}sica y Astronom\'{i}a, Facultad de Ciencias, Universidad de Valpara\'{i}so, Gran Bretana N 1111, Playa Ancha, Valpara\'{i}so, Chile}
\affil{$^{4}$EMBIGGEN Anillo, Concepci\'{o}n, Chile}
\affil{$^{5}$Center for Relativistic Astrophysics, School of Physics, Georgia Institute of Technology, Atlanta, GA 30332, USA}
\affil{$^{6}$Cahill Center for Astronomy and Astrophysics, California Institute of Technology, Pasadena, CA 91125, USA}
\affil{$^{7}$Pontificia Universidad Cat\'{o}lica de Chile, Instituto de Astrof\'{i}sica, Casilla 306, Santiago 22, Chile}
\affil{$^{8}$Millenium Institute of Astrophysics, Santiago, Chile}
\affil{$^{9}$Space Science Institute, 4750 Walnut Street, Suite 205, Boulder, Colorado 80301}
\affil{$^{10}$Space Sciences Laboratory, University of California, Berkeley CA 94720, USA}
\affil{$^{11}$Department of Astronomy and Astrophysics, The Pennsylvania State University, 525 Davey Lab, University Park, PA 16802, USA}
\affil{$^{12}$Institute for Gravitation and the Cosmos, The Pennsylvania State University, University Park, PA 16802, USA}
\affil{$^{13}$Department of Physics, The Pennsylvania State University, 525 Davey Lab, University Park, PA 16802, USA}
\affil{$^{14}$DTU Space, National Space Institute, Technical University of Denmark, Elektrovej 327, DK-2800 Lyngby, Denmark}
\affil{$^{15}$Lawrence Livermore National Laboratory, Livermore, CA 94550, USA}
\affil{$^{16}$Columbia Astrophysics Laboratory, Columbia University, New York, NY 10027, USA}
\affil{$^{17}$Department of Physics and Astronomy, Darmouth College, 6127 Wilder Laboratory, Hanover, NH 03755, USA}
\affil{$^{18}$Dipartimento di Matematica e Fisica, Universit\'{a} degli Studi Roma Tre, via della Vasca Navale 84, I-00146 Roma, Italy}
\affil{$^{19}$ASI Science Data Center, via Galileo Galilei, I-00044 Frascati, Italy}
\affil{$^{20}$Observatorio Astronomico di Roma (INAF), via Frascati 33, 00040 Monte Porzio Catone (Roma), Italy}
\affil{$^{21}$Department of Astronomy, Kyoto University, Kitashirakawa-Oiwake-cho, Sakyo-ku, Kyoto 606-8502, Japan}
\affil{$^{22}$NASA Goddard Space Flight Center, Greenbelt, MD 20771, USA}
\affil{$^{23}$Jet Propulsion Laboratory, California Institute of Technology, Pasadena, CA 91109, USA}

\begin{abstract}
We present two \textsl{NuSTAR} observations of the local Seyfert 2 active galactic nucleus (AGN) and an ultraluminous X-ray source (ULX) candidate in NGC 5643. Together with archival data from \textsl{Chandra}, \textsl{XMM-Newton} and \textsl{Swift}-BAT, we perform a high-quality broadband spectral analysis of the AGN over two decades in energy ($\sim$0.5--100 keV). Previous X-ray observations suggested that the AGN is obscured by a Compton-thick (CT) column of obscuring gas along our line-of-sight. However, the lack of high-quality $\gtrsim$ 10 keV observations, together with the presence of a nearby X-ray luminous source, NGC 5643 X--1, had left significant uncertainties in the characterization of the nuclear spectrum. \textsl{NuSTAR} now enables the AGN and NGC 5643 X--1 to be separately resolved above 10 keV for the first time and allows a direct measurement of the absorbing column density toward the nucleus. The new data show that the nucleus is indeed obscured by a CT column of $N_{\rm{H}}$ $\gtrsim$ 5 $\times$ 10$^{24}$ cm$^{-2}$. The range of 2--10 keV absorption-corrected luminosity inferred from the best fitting models is \emph{L}$_{2-10,\rm{int}} =$ (0.8--1.7) $\times$ 10$^{42}$ erg s$^{-1}$, consistent with that predicted from multiwavelength intrinsic luminosity indicators. We also study the \textsl{NuSTAR} data for NGC 5643 X--1, and show that it exhibits evidence for a spectral cut-off at energy, $E$ $\sim$ 10 keV, similar to that seen in other ULXs observed by \textsl{NuSTAR}. Along with the evidence for significant X-ray luminosity variations in the 3--8 keV band from 2003--2014, our results further strengthen the ULX classification of NGC 5643 X--1.
\end{abstract}



\section{Introduction}

Compton-thick AGN (CTAGN) are expected to constitute a significant fraction of the overall AGN population in the local universe, accounting for $\sim$20--30$\%$ of AGN according to multiwavelength studies (e.g. \citealp{Risaliti99}; \citealp{Burlon11}; \citealp{Goulding11}). Many studies also predict that CTAGN provide a substantial contribution to the cosmic X-ray background (CXB), responsible for 10--25$\%$ of the flux at the peak energy, $\sim$ 30 keV (e.g. \citealp{Gilli07}; \citealp{Treister09}; \citealp{DraperBallantyne10}; \citealp{Akylas12}; \citealp{Ueda14}). Yet, their census is still far from complete. The high line-of-sight column density in CTAGN ($N_{\rm{H}}$ $\geq$ 1/$\sigma_{T}$ $=$ 1.5 $\times$ 10$^{24}$ cm$^{-2}$, where $\sigma_{T}$ is the Thomson scattering constant), generally attributed to the parsec-scale circumnuclear torus of AGN unification schemes, as well as larger-scale molecular clouds and dust lanes, results in severe attenuation of the direct X-ray emission from CTAGN at energies below 10 keV. This is why observations at higher energies are needed to probe this direct component and provide unambiguous identification of CTAGN. However, even at $E$ $>$ 10 keV, the remaining flux that we observe in the most extreme CTAGN will be from photons scattered or reflected from the back-side of the torus, and will comprise just a few percent of the intrinsic AGN power (e.g. \citealp{Iwasawa97}; \citealp{Matt00}; \citealp{Balokovic14}). This makes the identification and characterization of CTAGN a challenging task.

To date, only $\sim$20 AGN within $\approx$ 200 Mpc have been confirmed as CT based upon detailed X-ray spectral characterization (\citealp{DellaCeca08}; \citealp{Goulding12}; \citealp{Gandhi14}). This corresponds to a fraction of $\ll$ 1$\%$ of the \emph{total} AGN population expected within that volume, suggesting that the vast majority of CTAGN are yet to be found even in the local universe. These \emph{bona-fide} CTAGN were unambiguously identified based upon a detection at energies above 10 keV and the presence of a Fe K$\alpha$ line at 6.4 keV with high equivalent width, EW $\gtrsim$ 1 keV. The identification and characterization of all the CTAGN is important in order to form an accurate census of accretion in the local universe, since much of the growth of supermassive black holes is thought to occur in such heavily obscured phases (e.g. \citealp{Fabian99}; \citealp{AlexanderHickox12}). Therefore, an accurate local benchmark is important for extrapolating the results to higher redshifts.

We have started a program to study a complete, volume-limited ($D$ $<$ 15 Mpc), mid-infrared (MIR) selected AGN sample from \citet{Goulding09}, with the main goal of constraining the population of CTAGN and the $N_{\rm{H}}$ distribution of AGN in the local universe. CTAGN candidates from the sample were identified using multiwavelength selections, such as X-ray spectroscopy, and intrinsic 2--10 keV luminosity indicators from high spatial resolution MIR 12$\mu$m continuum and [O{\sc{iii}}]$\lambda$5007\AA \ line luminosity corrected for the Balmer decrement. One of the candidates that stands out in the sample as being CT based on these analyses is NGC 5643. NGC 5643 is a nearby face-on (\emph{i} $\approx$ 30$^{\circ}$)\footnote{The host galaxy inclination was obtained from the HyperLeda website (http://leda.univ-lyon1.fr/).} SAB(rs)C galaxy hosting a low-luminosity Seyfert 2 nucleus (\citealp{Phillips83}). It has a redshift of $z$ $=$ 0.0040, corresponding to a metric/proper distance of $D$ $=$ 13.9 Mpc under the assumption of the \citet{Mould00} cosmic attractor model \citep{Sanders03}.\footnote{\citet{Mould00} adjusts heliocentric redshifts to the centroid of the Local Group, taking into account the gravitational attraction towards the Virgo Cluster, the Great Attractor and the Shapley supercluster.}

NGC 5643 features a compact radio core with two-sided, kiloparsec-scale lobes in an east-west orientation \citep{Morris85}, and a cospatial one-sided H$\alpha$ and [O{\sc{iii}}] emission line region extending east-ward of the nucleus for at least 1.8 kpc \citep{Simpson97}. Despite the intense star formation episodes occuring in the spiral arm, MIR diagnostics suggest that the AGN still dominates the overall IR (8--1000$\mu$m) energy budget \citep{Genzel98}. Comparisons of optical spectra with synthesis models, however, are consistent with a ``starburst/Seyfert 2 composite" spectrum \citep{CidFernandes01}. Using Br$\gamma$ emission, \citet{Davies14} found no on-going star formation activity in the nucleus, although the possibility of a recent (terminated) starburst cannot be excluded. This source also shows spatially resolved molecular gas flowing out from the AGN at a rate of 10 M$_{\odot}$ yr$^{-1}$ \citep{Davies14}. Although water maser emission associated with the AGN has been detected in NGC 5643, a corresponding spatially resolved map, which would allow for a direct measurement of the supermassive black hole mass (\emph{M}$_{\rm{BH}}$), is not yet available \citep{Greenhill03}. However, an indirect \emph{M}$_{\rm{BH}}$ measurement from the galaxy stellar velocity dispersion ($\sigma_{*}$) provides an estimated black hole mass of \emph{M}$_{\rm{BH}}$ $=$ 10$^{6.4}$ M$_{\odot}$ \citep{Goulding10}.\footnote{The same \emph{M}$_{\rm{BH}}$ estimate is obtained using an updated value of $\sigma_{*}$ derived from the [O {\sc{iii}}]$\lambda$5007\AA \ emission line width from \citet{Gu06} and the latest \emph{M}$_{\rm{BH}}$--$\sigma_{*}$ correlation by \citet{McconnellMa13}.}

In X-rays, NGC 5643 has been observed by \textsl{ASCA} \citep{Guainazzi04}, \textsl{BeppoSAX} \citep{Maiolino98}, \textsl{ROSAT} \citep{Guainazzi04}, \textsl{Chandra} \citep{Bianchi06} and \textsl{XMM-Newton} (\citealp{Guainazzi04}; \citealp{Matt13}). Dramatic spectral changes were observed between the \textsl{XMM-Newton} observation carried out in 2003, and the \textsl{ASCA} and \textsl{BeppoSAX} observations performed earlier. However, the point spread function (PSF) of \textsl{ASCA} and \textsl{BeppoSAX} were not sufficient to separate the emission of the nucleus from that of a nearby X-ray source (at an angular separation of $\sim$50$\arcsec$), NGC 5643 X--1, which was found to be very bright at the time of the \textsl{XMM-Newton} observation (G04). Therefore, it remained unclear which source was responsible for the spectral variability observed. Comparisons of the \textsl{XMM-Newton} observations in 2003 and 2009 showed that there is no significant variation in the spectrum of the AGN (\citealp{Matt13}; hereafter M13). However, the off-nuclear source was found to be more than a factor of two fainter in flux in 2009 (M13) than in 2003 (G04). 

The \textsl{Chandra} image of the AGN shows that the soft X-ray emission ($E$ $\lesssim$ 2 keV) of the nucleus is spatially correlated with the [O{\sc{iii}}] emission \citep{Bianchi06}, consistent with what is commonly observed for many Seyfert 2 galaxies. The dominant power source of this soft X-ray emission appears to be photoionization from the AGN, although it is still unclear how much collisional ionization contributes to the overall X-ray emission (M13). Above 2 keV, the X-ray spectra from these various observations show indications of the nucleus being absorbed by CT material. The evidence for this are the detection of a prominent Fe K$\alpha$ line (EW $>$ 1 keV) and a flat photon index below 10 keV ($\Gamma_{2-10}<$ 1), which are characteristics of a reflection-dominated spectrum. Analysis of the low signal-to-noise ratio (S/N) \textsl{BeppoSAX} spectrum where NGC 5643 was detected only up to 10 keV, combined with an upper limit for the 15--100 keV band, suggested a tentative lower limit to the column density of $N_{\rm{H}}$ $>$ 10$^{25}$ cm$^{-2}$ \citep{Maiolino98}. 

The true nature of the off-nuclear X-ray source, NGC 5643 X--1, is still uncertain. It is located in the outskirts of the host galaxy optical emission and identification of counterparts at other wavelengths has been ambiguous (G04). It is highly likely that the source is located inside the galaxy and therefore would be a powerful ultraluminous X-ray source (ULX) with \emph{L}$_{2-10}$ $\approx$ 1.7 $\times$ 10$^{40}$ erg s$^{-1}$ based on the flux observed in 2003 (G04). This is comparable to the observed luminosity of the AGN itself; i.e., \emph{L}$_{2-10}$ $\approx$ 1.9 $\times$ 10$^{40}$ erg s$^{-1}$ (G04).

\begin{figure*}
\epsscale{0.8}
\plotone{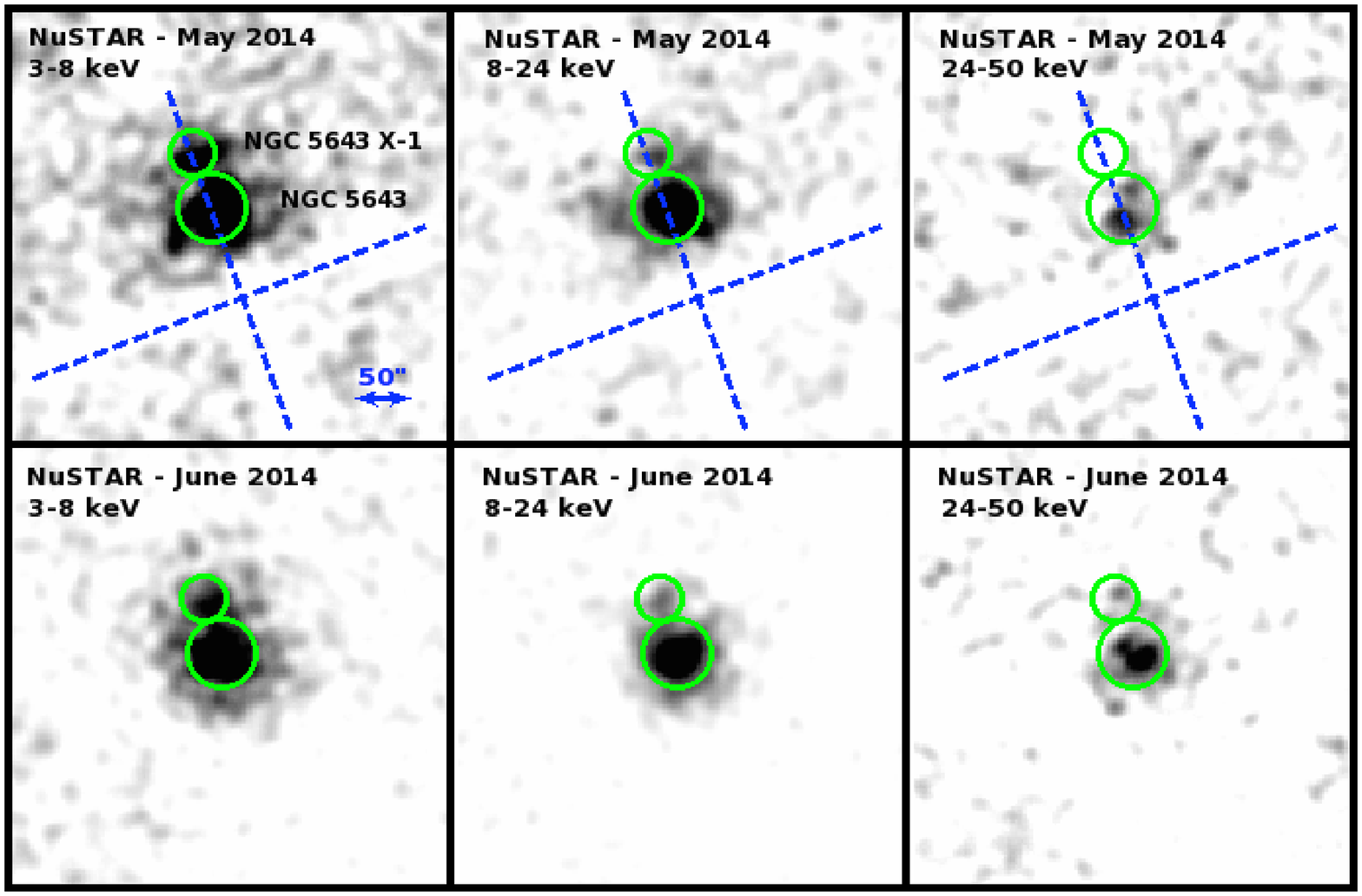}
\epsscale{0.5}
\plotone{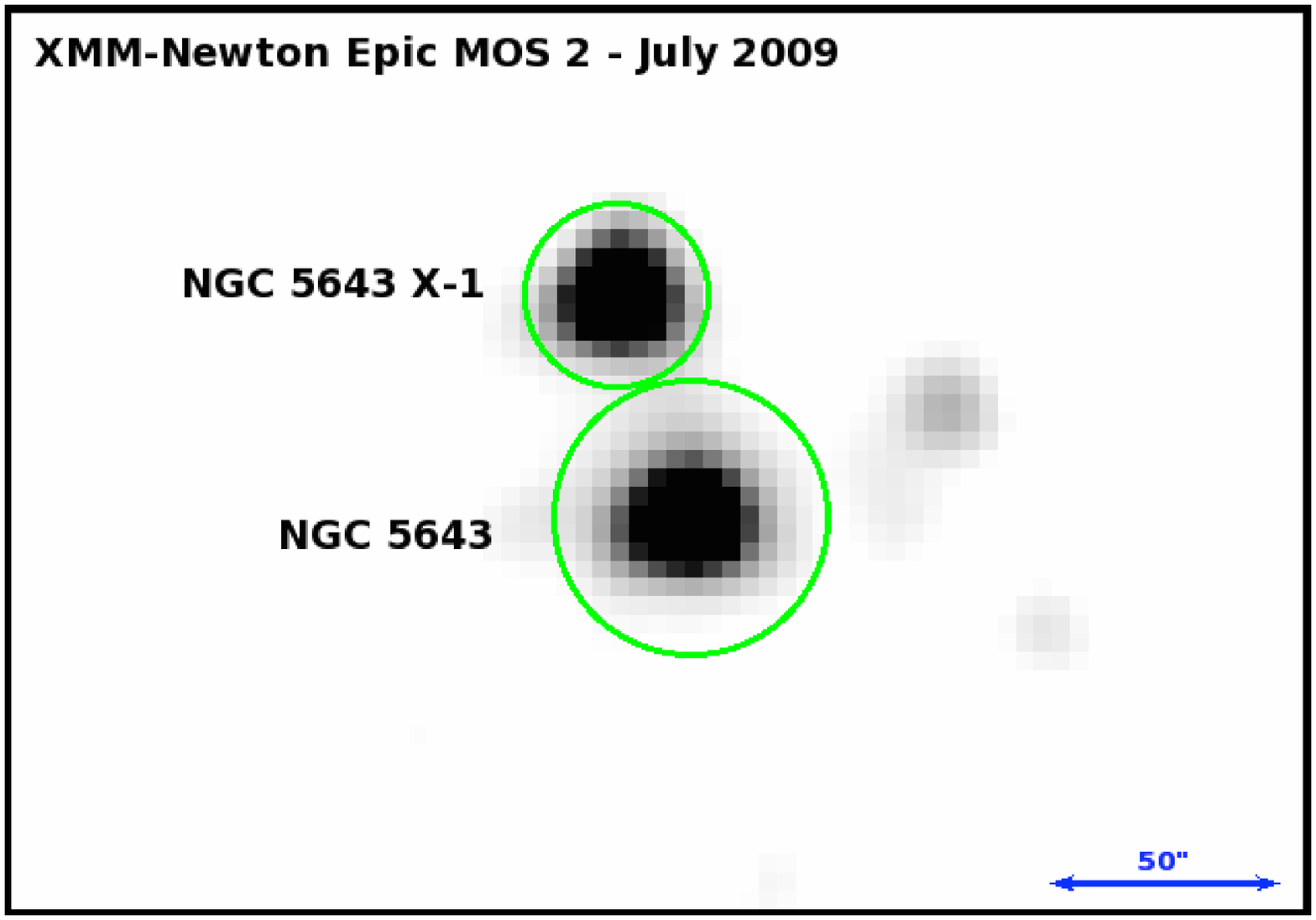}
\caption{\textsl{NuSTAR} and \textsl{XMM-Newton} images of the AGN (NGC 5643) and the ULX candidate (NGC 5643 X--1). The sources are circled in green with 30$\arcsec$ and 20$\arcsec$-radius apertures, respectively. Images are smoothed with a Gaussian function of radius 5 and 3 pixels, corresponding to 12.3$\arcsec$ and 3.3$\arcsec$ for \textsl{NuSTAR} and \textsl{XMM-Newton}, respectively. North is up and East is to the left in all images. Top: \textsl{NuSTAR} combined FPMA+B images in the 3--8, 8--24 and 24--50 keV band for the May 2014 observation (first row) and June 2014 observation (second row). Blue dashed lines correspond to the gap between the \textsl{NuSTAR} detectors for the May observation. Bottom: \textsl{XMM-Newton} EPIC MOS 2 image (July 2009) in the 0.5--10 keV band shown to provide a clearer view of the two sources at lower energies.}
\end{figure*}

\begin{table*}
\begin{center}
\caption{X-ray Observations log of NGC 5643\label{table1}}
\begin{tabular}{lccccc}
\tableline
 \ \ \ \ \ \ \ \ \ Instrument & ObsID & Date & Energy Band & Net Exposure time & Net count Rate \\
          &       &      &    (keV)       & (ks)      & (10$^{-2}$ cts s$^{-1}$) \\
 \ \ \ \ \ \ \ \ \ \ \ \ \ \        (1) & (2) & (3) & (4) & (5) & (6) \\
\tableline
\textsl{Chandra} ACIS-S & 5636 & 2004-12-26 & 0.5--8 & 7.63 & 4.96 \\
\textsl{XMM-Newton} PN/MOS1+2 & 0601420101 & 2009-07-25 & 0.5--10 & 45.4/53.4 & 14.6/3.76 \\
\textsl{Swift}-BAT & - & 2004-2010 & 14--100 & 7340 & 0.00223  \\
\textsl{Swift}-XRT & 00080731001 & 2014-05-24 & 0.5--10 & 3.96 & 0.472 \\
\textsl{NuSTAR} FPMA/FPMB & 60061362002 & 2014-05-24 & 3--50 & 22.5/22.4 & 2.44/1.94 \\
\textsl{NuSTAR} FPMA/FPMB & 60061362004 & 2014-06-30 & 3--50 & 19.7/19.7 & 2.32/2.36 \\
\tableline
\end{tabular}
\tablecomments{(1): List of observatories and instruments; (2): Observation identification number (obsID); (3): Observation UT start date; (4) Energy band in keV; (5) The net exposure time in ks; (6): Net count rate for the AGN in the given energy band in units of 10$^{-2}$cts s$^{-1}$.}     
\end{center}
\end{table*}

In this paper, we present new \textsl{NuSTAR} observations of NGC 5643 in which the AGN and the off-nuclear source, NGC 5643 X--1, are clearly resolved and detected at hard X-ray energies ($>$ 10 keV) for the first time. This allows us to provide the most accurate spectral analysis of the AGN to date. The aim of this paper is to characterize the broadband spectrum of the AGN by combining our \textsl{NuSTAR} data with existing data from \textsl{Chandra}, \textsl{XMM-Newton} and \textsl{Swift}-BAT. We also present the \textsl{NuSTAR} data for NGC 5643 X-1, which is detected above 10 keV for the first time.

The paper is organized as follows: we describe details of the X-ray observations and data reduction of the AGN in Section 2, followed by the spectral fitting procedures and results in Section 3. In Section 4, we present the data analysis and results on NGC 5643 X--1. This is followed by a discussion in Section 5. The paper concludes with a summary in Section 6. 


\section{Observations}

In this section, we describe the \textsl{NuSTAR} observations and data analysis procedures for the AGN. We also detail the archival \textsl{Chandra}, \textsl{XMM-Newton} and \textsl{Swift} data that were used to facilitate our broadband spectral analysis. Details of these observations are provided in Table 1, and the data reduction is described below. 

\subsection{\textsl{NuSTAR}}
\textsl{NuSTAR} (\textsl{Nuclear Spectroscopic Telescope Array}; \citealp{Harrison13}), operating at 3--79 keV, is the first orbiting observatory with the ability to focus X-ray photons at $E$ $>$ 10 keV. It provides a two orders of magnitude improvement in sensitivity and over an order of magnitude improvement in angular resolution with respect to previous hard X-ray orbiting observatories. It consists of two co-aligned focal plane modules, FPMA and FPMB, each covering the same 12$\arcmin$ $\times$ 12$\arcmin$ portion of the sky, and each comprised of four detectors placed in a 2 $\times$ 2 array. \textsl{NuSTAR} has an angular resolution of 18$\arcsec$ FWHM with a half power diameter of $\sim$1$\arcmin$, and energy resolutions (FWHM) of 0.4 and 0.9 keV at 6 and 60 keV, respectively. These capabilities make \textsl{NuSTAR} an ideal instrument to characterize the spectral shape of heavily obscured AGN in the local universe (e.g. \citealp{Balokovic14}; \citealp{Gandhi14}; \citealp{Puccetti14}; \citealp{Bauer14}). 

NGC 5643 was observed twice by \textsl{NuSTAR} in 2014 -- an initial observation with a nominal exposure time of 22.5 ks taken in May, followed by an additional 19.7 ks observation in June. The second observation was conducted to improve the S/N of NGC 5643 as it fell near the detector gap in the first observation. The data were processed with the \textsl{NuSTAR} Data Analysis Software ({\sc{nustardas}}) v1.4.1 within {\sc{heasoft}} v6.15.1 with CALDB v20150316. Calibrated and cleaned event files were produced using the {\sc{nupipeline}} v0.4.3 script with standard filter flags. Spectra and response files were extracted using the {\sc{nuproducts}} v0.2.5 task. 

For each observation, the AGN spectra were extracted using a circular aperture region of 30$\arcsec$-radius centered on its peak emission. The background photon were collected from polygon-shaped regions lying on the same detector as the source for the June observation, and from two adjacent detectors for the May observation (since the source fell close to the detector gap in this observation). We excluded background photons that lay within a circular region of $\sim$ 70$\arcsec$-radius around the source to exclude emission from NGC 5643 X--1. Significant counts are detected up to 50 keV for the AGN. We show the combined FPMA+B images of the AGN and NGC 5643 X--1 in the 3--8, 8--24 and 24--50 keV bands in Figure 1. 

\subsection{\textsl{XMM-Newton}}

NGC 5643 was also observed on two occasions by \textsl{XMM-Newton}, for 10 ks in 2003 and 55 ks in 2009. In both cases, the observations were performed with the EPIC CCD cameras (PN, MOS1, MOS2), operated in full frame mode with the medium filter. The observations were discussed in detail in G04 and M13, respectively. For the AGN, we only extracted and used the spectra from the longer 2009 observation as it has the highest S/N. Data were reduced within {\sc{sas}} v1.2, screened for flaring particle background as described in M13. 

Source spectra were extracted within a 25$\arcsec$-radius aperture centered on the AGN for all cameras. Background photons were selected from source-free circular regions of 100$\arcsec$-radius on the same chip as the source. Patterns 0 to 4 and 0 to 12 were used for the PN and MOS spectra, respectively. The spectra from MOS1 and MOS2 were co-added as the data were consistent with each other (see Section 3 for further details).

\subsection{\textsl{Chandra}}

NGC 5643 has only been observed once by \textsl{Chandra}. This observation was conducted in 2004 with the ACIS-S detector with an exposure time of $\sim$ 8 ks. The results of the observation were first published in \citet{Bianchi06}. We reprocessed the data to create event files with updated calibration modifications using the CIAO v4.6 pipeline following standard procedures. 

Source counts were extracted using the {\sc{specextract}} task in CIAO from a circular region of 25$\arcsec$-radius centered on the AGN to match the \textsl{XMM-Newton} extraction region. The background was extracted from a source-free 30$\arcsec$-radius circular region close to the source. 

\subsection{\textsl{Swift}}

The first \textsl{NuSTAR} observation in May 2014 was accompanied by a $\sim$ 5 ks \textsl{Swift} X-ray Telescope (XRT; \citealp{Burrows05}) observation, which started approximately an hour after the \textsl{NuSTAR} observation began. The purpose of this observation was to provide simultaneous coverage for the soft X-ray end of the spectrum ($\lesssim$ 3 keV) where the \textsl{NuSTAR} sensitivity drops off. The data were reduced using the {\sc{xrtpipeline}} v0.13.0, which is part of the XRT Data Analysis Software ({\sc{xrt-das}}) within {\sc{heasoft}}. However, with only $\sim$ 20 counts between 0.5--10 keV, the exposure is not long enough to provide additional constraints beyond those already obtained with \textsl{NuSTAR}, \textsl{XMM-Newton} and \textsl{Chandra}. Therefore, we only used these data to check for consistency with the \textsl{XMM-Newton} and \textsl{Chandra} data at 0.5--10 keV (see Section 3).

The Burst Alert Telescope (BAT; \citealp{Barthelmy05}) onboard \textsl{Swift} has been continuously monitoring the sky at 14--195 keV, producing images of a large number of hard X-ray sources thanks to its wide field of view and large sky coverage. We used the stacked 70-month spectrum and its associated response file downloaded from the \textsl{Swift}-BAT 70-month Hard X-ray Survey Source Catalog \citep{Baumgartner13} to provide X-ray constraints above the \textsl{NuSTAR} band.\footnote{The \textsl{Swift}-BAT 70-month Hard X-ray Survey Source Catalog is available online at http://swift.gsfc.nasa.gov/results/bs70mon/} NGC 5643 is detected at the 5.4$\sigma$ significance level with significant counts up to $\sim$ 100 keV.


\section{Broadband Spectral Modeling of the AGN}

\begin{figure*}
\epsscale{.58}
\plotone{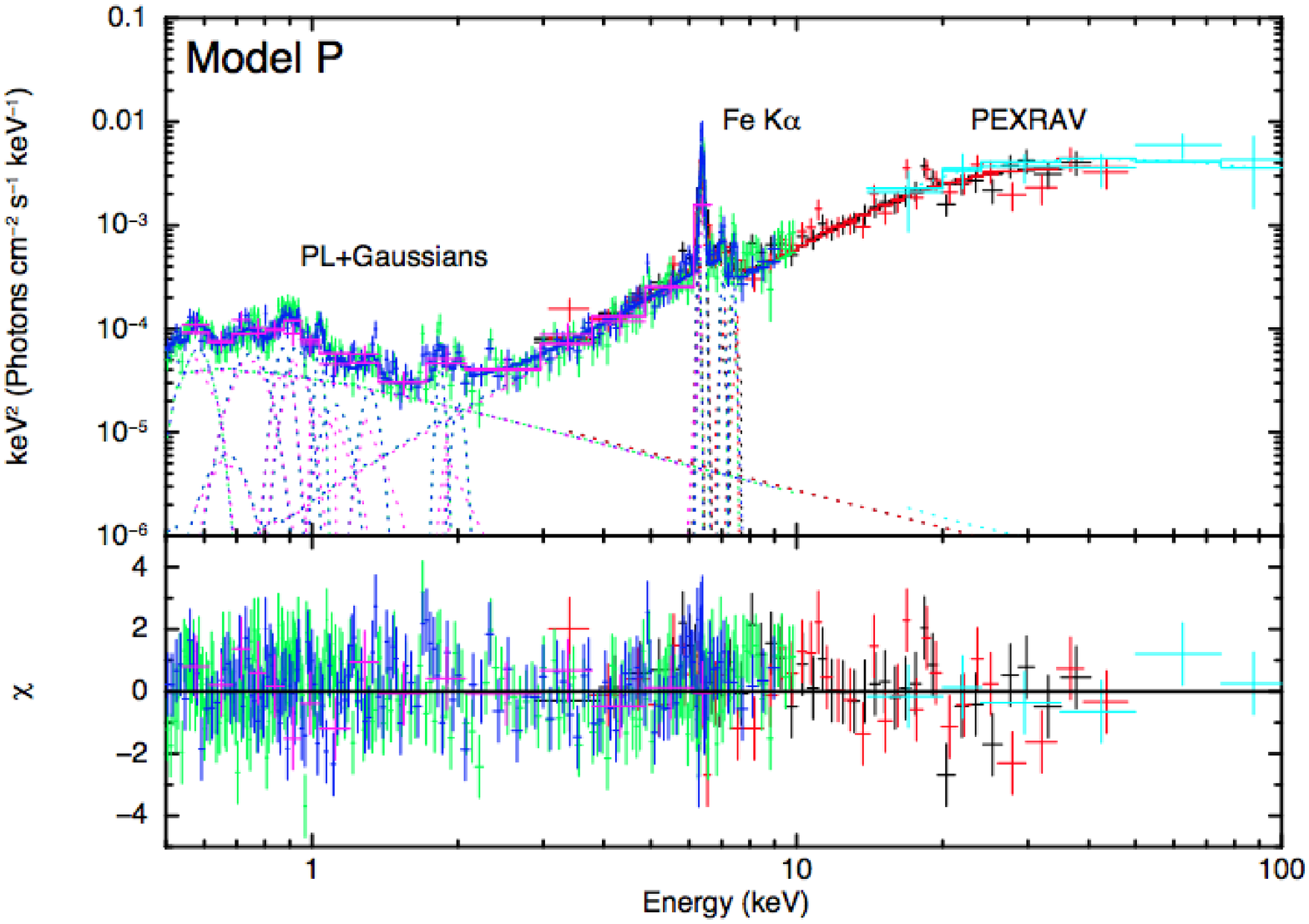}
\plotone{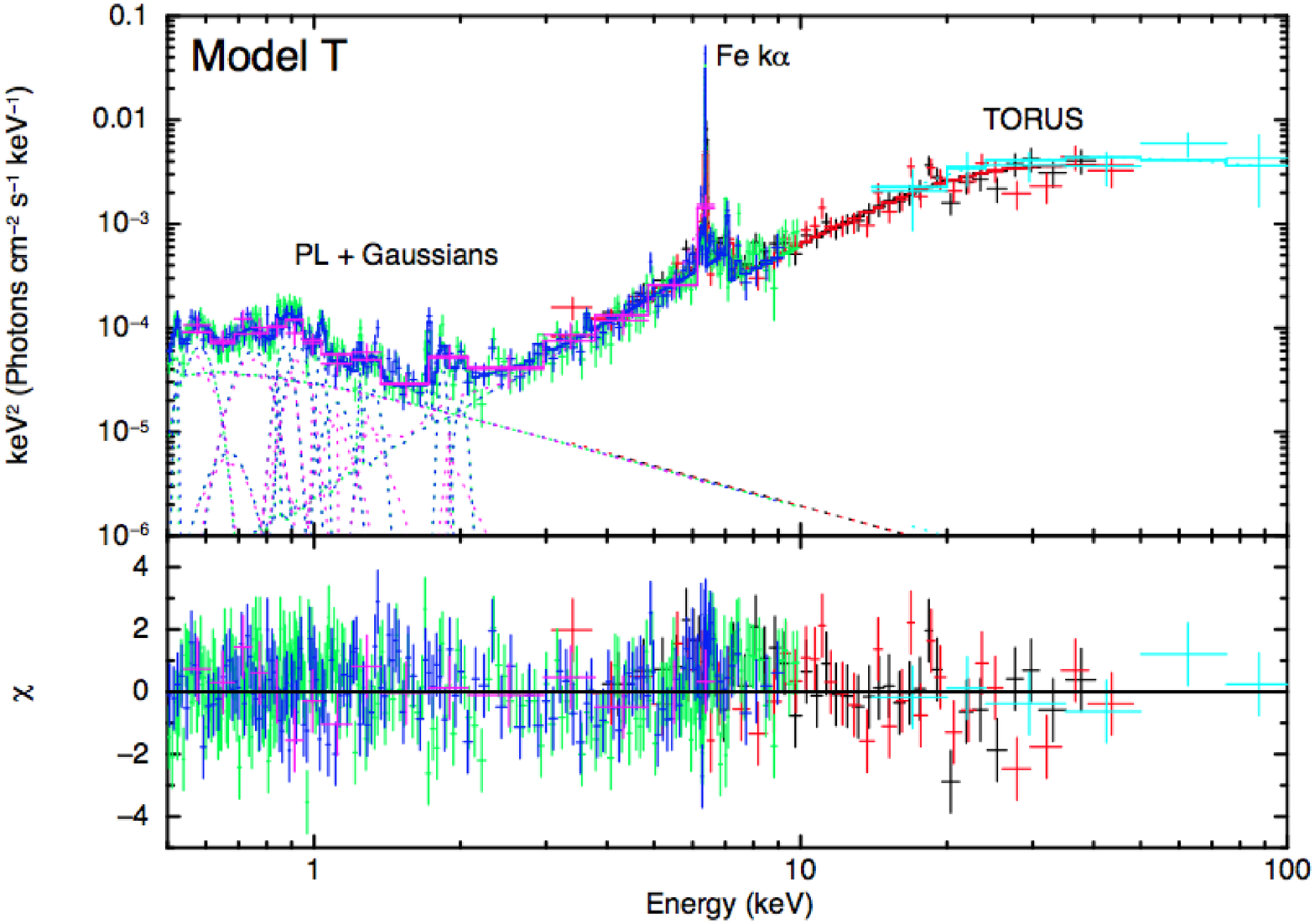}
\plotone{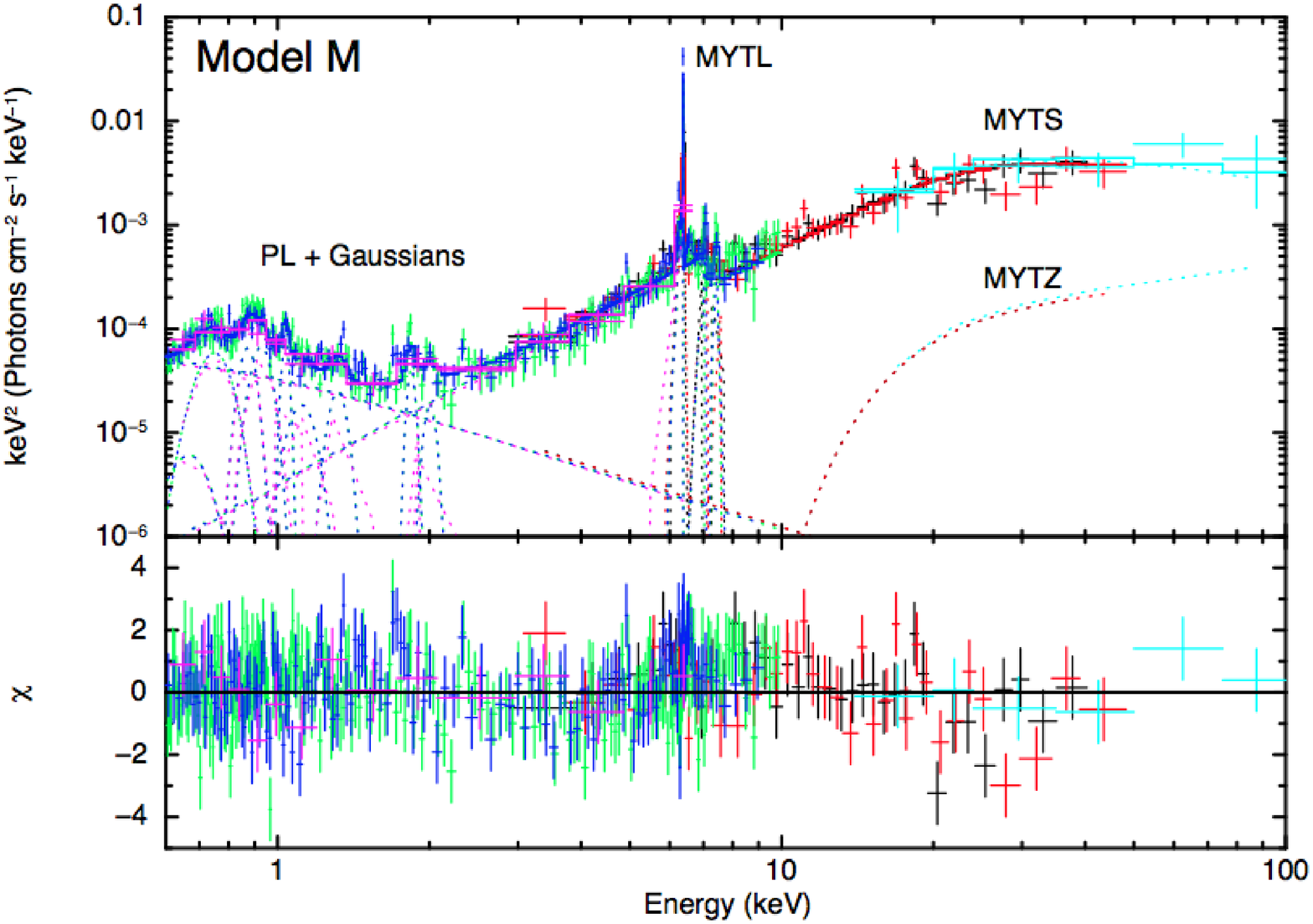}
\caption{Best-fit models to the combined \textsl{NuSTAR}, \textsl{Chandra}, \textsl{XMM-Newton} and \textsl{Swift}-BAT data of the AGN - Model P (top), Model T (bottom left), and Model M (bottom right). Model P and T were fitted between 0.5--100 keV, and Model M was fitted between 0.6--100 keV since we found strong residuals near $\sim$0.5 keV for this model. The lower energy data ($\lesssim$ 2 keV) were modeled using a soft power-law (PL) and Gaussian lines to simulate photoionization by the AGN. The top panels of each plot show the unfolded model in $EF_{E}$ units, while the bottom panels show the fit residuals in terms of sigma with error bars of size one. Color scheme: black (\textsl{NuSTAR} FPMA), red (\textsl{NuSTAR} FPMB), blue (\textsl{XMM-Newton} PN), green (\textsl{XMM-Newton} MOS), purple (\textsl{Chandra}), cyan (\textsl{Swift}-BAT).}
\end{figure*}

\begin{table*}
\begin{center}
\caption{X-ray spectral fitting results for the AGN in NGC 5643.}
\begin{tabular}{lccccc}
\tableline
Component & Parameter & Units & Model P & Model T & Model M \\
\tableline
Absorber/Reflector & $N_{\rm{H}}$(eq) & 10$^{24}$ cm$^{-2}$ & - & 100$^{+u}_{-85}$ & 10.0$^{+u}_{-0.5}$ \\
                   & $N_{\rm{H}}$(los) & 10$^{24}$ cm$^{-2}$ & - &100$^{+u}_{-85}$ & 5.8$^{+u}_{-1.2}$ \\
                   & $\theta_{\rm{inc}}$ &  deg & 65$^{f}$ & 65$^{f}$ & 65.9$^{+2.3}_{-1.9}$  \\
                   & $\theta_{\rm{tor}}$ &  deg & - & 60.0$^{+0.01}_{-1.33}$ & -  \\
AGN Continuum      & $\Gamma_{\rm{soft}}$ & &  3.19$\pm$0.13 & 3.28$^{+0.14}_{-0.12}$ & 3.54$^{+0.16}_{-0.15}$  \\
                   & $\Gamma_{\rm{hard}}$ & & 1.79$\pm$0.05 & 1.97$^{+0.03}_{-0.05}$ & 2.10$^{+0.04}_{-0.03}$  \\
                   & \emph{L}$_{0.5-2\rm{,obs}}$ & 10$^{40}$ erg s$^{-1}$ & 0.4 &0.4 & 0.4 \\
                   & \emph{L}$_{2-10\rm{,obs}}$ & 10$^{40}$ erg s$^{-1}$ & 1.7 & 1.7 & 1.6  \\
                   & \emph{L}$_{30-100\rm{,obs}}$ &  10$^{40}$ erg s$^{-1}$ & 15.9 & 16.2 & 16.3 \\
                   & \emph{L}$_{0.5-2\rm{,int}}$ & 10$^{42}$ erg s$^{-1}$ & 0.4 &  0.6 & 0.3 \\
                   & \emph{L}$_{2-10\rm{,int}}$ & 10$^{42}$ erg s$^{-1}$ & 1.7 & 0.8 & 0.9\\
                   & \emph{L}$_{30-100\rm{,int}}$ &  10$^{42}$ erg s$^{-1}$ & 1.6 & 0.8 & 0.6 \\
\textsl{C}$^{\rm{XMM}}_{\rm{NuSTAR}}$ &&  & 0.94$\pm$0.05 & 0.97$^{+0.06}_{-0.04}$  & 0.94$\pm$0.05 \\
\textsl{C}$^{\rm{Chandra}}_{\rm{NuSTAR}}$ & & & 0.91$\pm$0.10 & 0.94$^{+0.11}_{-0.09}$  & 0.90$_{-0.05}^{+0.10}$ \\
\textsl{C}$^{\rm{BAT}}_{\rm{NuSTAR}}$ & & & 1.23$^{+0.32}_{-0.31}$ & 1.20$\pm$0.30 & 1.15$^{+0.28}_{-0.29}$ \\
$\chi^{2}_{\rm{r}}$/d.o.f. & & & 1.10/500  & 1.16/499 & 1.21/471 \\           
\tableline
\end{tabular}
\tablecomments{$^{f}$fixed, $^{u}$unconstrained. Best-fitting model parameters for Model P ({\sc{pexrav}} model by \citealp{MZ95}), T ({\sc{torus}} model by \citealp{BN11}) and M ({MYT\sc{orus}} model by \citealp{MY09}). Details of each model are described in Section 3.} 
\end{center}
\end{table*}

We describe the broadband X-ray spectral analysis of the AGN in this section. The analysis was carried out using {\sc{xspec}} v12.8.2.\footnote{The {\sc{XSPEC}} manual can be downloaded from http://heasarc.gsfc.nasa.gov/xanadu/xspec/XspecManual.pdf} All spectra were binned to a minimum of 25 counts per bin to allow the use of $\chi^{2}$ statistics. Absorption through a fixed Galactic column along the line-of-sight, $N_{\rm{H}}^{\rm{Gal}}$ $=$ 8.01 $\times$ 10$^{20}$ cm$^{-2}$ \citep{Kalberla05}, was included in all spectral fits using the {\sc{xspec}} model ``{\sc{phabs}}", and solar abundances were assumed for all models. All errors are quoted at 90$\%$ confidence, unless stated otherwise. Details of the main results are summarized in Table 2, and the best fit spectra are shown in Figure 2.

We began our spectral modelling with an examination of the \textsl{NuSTAR} data alone. We started by modeling the spectra of each of the \textsl{NuSTAR} observations in the 3--50 keV band using a simple absorbed power-law model with the column density fixed to the Galactic value. A prominent excess of counts just above 6 keV was observed in both spectra, suggesting the presence of Fe K$\alpha$ emission. The parameters returned by the two spectra were consistent with each other, indicating that there are no significant differences between the two observations. Therefore, we co-added the spectra for each FPM using the {\sc{addascaspec}} script (the same test and procedures were also done for the two \textsl{XMM-Newton} MOS spectra). 

We then modeled the combined \textsl{NuSTAR} spectra in the 3--50 keV band with an absorbed power-law model (column density fixed to $N_{\rm{H}}^{\rm{Gal}}$) and a Gaussian line at $E$ $\approx$ 6.4 keV. The model measured a photon index and Fe K$\alpha$ EW of $\Gamma$ $=$ 0.55 $\pm$ 0.07 and EW$_{\rm{Fe}}$ $=$ 2.22$_{-0.34}^{+0.35}$ keV, respectively, with a fit statistic of $\chi^{2}$ = 115 for 76 degrees of freedom (d.o.f). The line energy is centered at \emph{E} $=$ 6.36 $\pm$ 0.04 (statistical)  $\pm$ 0.04 (systematic; \citealp{Madsen15}) keV, consistent with neutral Fe K$\alpha$ emission. The very flat $\Gamma$ and large EW$_{\rm{Fe}}$, which were also found in previous observations, are characteristic signatures of Compton-thick absorption and reflection from optically thick cold gas. 

We checked for variability in the flux of the AGN in the 0.5--10 keV band by comparing the \textsl{XMM-Newton} and \textsl{Chandra} data with the \textsl{Swift}-XRT data. The fluxes of all spectra are consistent with each other, \emph{f}$_{0.5-10.0}$ $\sim$ 10$^{-12}$ erg s$^{-1}$ cm$^{-2}$, suggesting that there are no significant differences between the three observations. We also checked for variability at hard X-ray energies by constructing power spectra from the 3--50 keV \textsl{NuSTAR} lightcurves from the two observations. These were then compared to the power spectra expected for pure Poisson noise and for the expected transmitted component variability observed in an unobscured AGN with a similar black hole mass and accretion rate (see \citealp{Arevalo14} for further details). No evidence for variability was found, however, given the low count rates, this test is not very sensitive. Given that the \textsl{NuSTAR} spectra from the two observations are consistent with each other, we assumed that the AGN has not varied. 

We therefore proceeded to fit the \textsl{XMM-Newton} and \textsl{Chandra} spectra combined with the \textsl{NuSTAR} and \textsl{Swif}-BAT spectra using more detailed physical models to better characterize the broadband spectrum of the AGN. We describe the details and results of each model used: Model P ({\sc{pexrav}} model by \citealp{MZ95}), T ({\sc{torus}} model by \citealp{BN11}) and M ({MYT\sc{orus}} model by \citealp{MY09}), in Section 3.1, 3.2 and 3.3, respectively. In addition to these models, we added extra components required to provide a good fit to the data, briefly described below. 

The soft energy ($\lesssim$ 2 keV) part of the spectra covered by \textsl{XMM-Newton} and \textsl{Chandra} is dominated by emission from photoionized material (M13). We modeled this emission based on M13 using a soft power-law\footnote{The soft power-law used to model the unresolved $\lesssim$ 2 keV emission also includes the scattered power-law component which is commonly used to simulate the AGN emission scattered into our-line-of sight by diffuse hot gas.} and 10 Gaussian components (9 for Model M; see Section 3.3) to model the emission lines (refer to Table 1 in M13). We also added four and one more Gaussian component(s) to Model P and M, respectively, to model the fluorescence lines emitted at $E$ $\gtrsim$ 2 keV which are not included in these models (see Section 3.1 and 3.3 respectively). 

Cross-calibration uncertainties between each observatory with respect to \textsl{NuSTAR} were included as free parameters ({\sc{constant}}, \textsl{C}). Initially, we left both the \textsl{XMM-Newton} EPIC-MOS and PN cameras constants free to vary. However, we found that their values are consistent with each other and decided to tie them together in the final fittings of all models. 
\\
\\
Our three models can be described as follows:\\
\\
Model P $=$ {\sc{constant}} $\times$ {\sc{phabs}} $\times$ ({\sc{pow}} $+$ \\ \\ {\sc{pexrav}} $+$ 14 $\times$ {\sc{zgauss}}),\ \ \ \ \ \ \ \ \ \ \ \ \ \ \ \ \ \ \ \ \ \ \ \ \ \ \ \ \  \ \ \ \ \ \ \ \ \ (1)
\\
\\
\\
Model T $=$ {\sc{constant}} $\times$ {\sc{phabs}} $\times$ ({\sc{pow}} $+$ \\ \\ {\sc{torus}} $+$ 10 $\times$ {\sc{zgauss}}). \ \ \ \ \ \ \ \ \ \ \ \ \ \ \ \ \ \ \ \ \ \ \ \ \ \ \ \ \ \ \ \ \ \ \ \ \ \ \ \ \ \ \ \ \ \ \ \ (2)
\\
\\
\\
Model M $=$ {\sc{constant}} $\times$ {\sc{phabs}} $\times$ ({\sc{pow}} $+$ \\ \\ {\sc{zpow}} $\times$ {\sc{mytz}} $+$ {\sc{myts}} $+$ {\sc{mytl}} $+$ \\ \\ 10 $\times$ {\sc{zgauss}}), \ \ \ \ \ \ \ \ \ \ \ \ \ \ \ \ \ \ \ \ \ \ \ \ \ \ \ \ \ \ \ \ \ \ \ \ \ \ \ \ \  \ \ \ \ \ \ \ \ \  \ \ \ \ \ \ \ \ \ (3)
\\
\\
\subsection{Model P}

In our fitting of the AGN broadband spectrum, we first consider the {\sc{pexrav}} reflection model (`Model P'; \citealp{MZ95}). This model has commonly been used to model reflection-dominated spectra. However, it assumes reflection off a slab geometry with an infinite column density, which is unlikely to represent the true geometry of the AGN torus. It also does not incorporate fluoerescence emission lines expected from a CTAGN such as the Fe K$\alpha$ and K$\beta$ lines. Because of these limitations, we also fitted the spectra of the AGN using more physically motivated reflection/obscuration models as described in Section 3.2 and 3.3. However, Model P can provide useful comparison with the other reflection models and previous studies of this source, as well as other CT sources.

We started by fixing all the line energies and normalizations of the soft Gaussian components to the values presented in M13. We also fixed the redshift at \emph{z} $=$ 0.0040. Initially, we set the intrinsic line widths to $\sigma$ $=$ 50 eV, as we expect them to be unresolved. We then allowed the widths for each line to vary in turn, to improve the fit. For those that returned values deviating significantly from 50 eV, we fixed them to this new value. 

We added several more Gaussian components to account for the Fe K$\alpha$ and Fe K$\beta$ lines, the Ni K$\alpha$ line and also the Compton shoulder associated with the Fe K$\alpha$ line which are not included in the model.\footnote{We note that modeling the Compton shoulder using a Gaussian component is incorrect since it is not exactly Gaussian, but we adopted this approach as an approximate parameterization for this particular model.} The parameters were also fixed to values obtained in M13, except for the line energy and normalization of the Fe K$\alpha$ line which were allowed to vary throughout. We fixed the inclination of the reflector to $\theta_{\rm{inc}}$ = 65$^{\circ}$, which is equal to that modeled by \citet{Fischer13} for NGC 5643 (with an uncertainty of $\pm$5$^{\circ}$) based on mapping the kinematics of the H$\alpha$ and [O{\sc{iii}}] narrow line region (NLR) emission observed by the \textsl{Hubble Space Telescope} (\textsl{HST}). The reflection scaling factor was fixed to \emph{R} $=$ $-$1 to simulate a reflection-dominated spectrum. Model P provides a good fit to the data ($\chi^{2}$/d.o.f $=$ 551/500), indicating that the spectra can be well-fitted by a pure reflection model without the need for any direct component.

The cross-calibration constants of \textsl{XMM-Newton}, \textsl{Chandra} and \textsl{Swift}-BAT relative to \textsl{NuSTAR} are consistent with each other with $C$ $\sim$ 1 within the statistical errors (see Table 2), indicating that the spectra are in good agreement with each other (see also \citealp{Madsen15} for the current cross-calibration status between different X-ray observatories and instruments with respect to \textsl{NuSTAR}). We estimated the intrinsic X-ray luminosities from this model by assuming that the observed luminosities at 0.5--10 keV and 30--100 keV are about 1$\%$ and 10$\%$ of the intrinsic luminosities at these energy bands, respectively (see Table 2). These inferred fractions were taken from detailed studies of NGC 1068 and have been widely used to estimate the intrinsic luminosities of other reflection-dominated AGN (e.g. \citealp{Matt97}; \citealp{Balokovic14}).
\\
\subsection{Model T}

We then proceeded to fit the AGN broadband spectrum using more physically motivated models, starting with the {\sc{torus}} model by \citet{BN11} (Model T). This model simulates obscuration by a medium with a conical section cut from a sphere, with a variable biconical polar opening angle ($\theta_{\rm{tor}}$) ranging between 26--84$^{\circ}$. Small values of $\theta_{\rm{tor}}$ correspond to a geometrically thick torus, while large values indicate a geometrically thin torus. The line-of-sight $N_{\rm{H}}$(los) through the torus, which is equal to the equatorial column density $N_{\rm{H}}$(eq), extends up to 10$^{26}$ cm$^{-2}$, allowing investigations of the most extreme obscuration, and is defined such that it is independent of the inclination angle ($\theta_{\rm{inc}}$). The model is defined between 0.1--320 keV and self-consistently predicts lines commonly found in obscured AGN, such as Fe K$\alpha$, Fe K$\beta$ and also K$\alpha$ emission from several other elements (C, O, Ne, Mg, Si, Ar, Ca, Cr, Fe and Ni). 

The soft components of the spectrum ($\lesssim$ 2 keV) were modeled as described earlier in Section 3 and 3.1. Since $\theta_{\rm{inc}}$ and $\theta_{\rm{tor}}$ could not be constrained simultaneously, we fixed $\theta_{\rm{inc}}$ to 65$^{\circ}$ \citep{Fischer13}. This model also yields a good fit to the data ($\chi^{2}$/d.o.f $=$ 579/499) with most fit residuals lying at the soft energies ($\lesssim$ 2 keV) and around the iron line emission ($\sim$6--7.5 keV; e.g. $\chi^{2}$/d.o.f $=$ 447/417 when fitting the model by ignoring the data around the iron line emission). The fitted intrinsic continuum power-law emission from this model has a photon index of $\Gamma$ $=$ 1.97$^{+0.03}_{-0.05}$ and is absorbed by a column density of $N_{\rm{H}}$(los) $>$ 1.5 $\times$ 10$^{25}$ cm$^{-2}$; i.e., heavily Compton-thick. The upper limit is unconstrained, with values of up to 10$^{26}$ cm$^{-2}$ allowed by the model. We found that the best-fit $\theta_{\rm{tor}}$ measured by this model is $\approx$ 60$^{\circ}$. As with Model P, the cross-calibration constants of each observation with respect to \textsl{NuSTAR} are consistent with 1. We inferred the intrinsic luminosities from this model based upon the best-fit parameters obtained, as presented in Table 2.  

We also tried to fit the spectrum using different $\theta_{\rm{inc}}$ to see how the key parameters would change. Setting $\theta_{\rm{inc}}$ to a lower value of 45$^{\circ}$ produced an unacceptable fit ($\chi^{2}$/d.o.f $=$ 3989/499). Fixing $\theta_{\rm{inc}}$  to a higher value of 85$^{\circ}$ to simulate a near edge-on torus inclination, however, provides a slight improvement to the fit ($\chi^{2}$/d.o.f $=$ 549/499). The column density obtained is consistent with what was inferred before, but is more constrained, $N_{\rm{H}}$(los) $=$ 1.8$_{-0.6}^{+4.3}$ $\times$ 10$^{25}$ cm$^{-2}$. The photon index is slightly lower,  $\Gamma$ $=$ 1.75$_{-0.03}^{+0.01}$, and the opening angle of the torus suggests a surprisingly thin torus, $\theta_{\rm{tor}}$  $=$ 78.5$_{-0.74}^{+0.01}$ degrees. However, we note that \citet{LiuLi15} claim that for an edge-on torus inclination, this model overestimates the reflection component, and therefore the parameters obtained may not be reliable. We thus favor the best-fitting model solution using $\theta_{\rm{inc}}$ $=$ 65$^{\circ}$.

A closer look at the fitted spectrum between 6--7.5 keV (Figure 3) indicates that the model overpredicts the data at $\approx$6.3 keV, which encompass the Compton shoulder associated with the Fe K$\alpha$ line emission, and underpredicts the data at $\approx$6.4--6.8 keV, which include the Fe K$\alpha$ line emission. We investigated whether the observed residuals could be caused by the presence of emission lines at 6.70 and 6.96 keV corresponding to ionized iron by adding Gaussian components at these energies. However, we found that the normalizations of the components are consistent with zero. We also tried to add a line smoothing component to the model ({\sc{gsmooth}}; energy index, $\alpha=$ 1) to see if the statistics could be further improved, but the fit indicated that this is not required ($\sigma_{{\rm{gsmooth}}}$ $\lesssim$ 40 eV). We noticed that the residuals are dominated by the \textsl{XMM-Newton} data, and therefore checked whether these could be caused by a shift in the energy scale of the data. Indeed we found that, applying an $\sim +11$ eV offset in the PN data can improve the overall quality of the fit ($\chi^{2}$/d.o.f $=$ 551/497) and diminish the residuals around the iron line complex (see also \citealp{Bauer14} for similar finding). However, this energy shift is consistent with the current calibration uncertainty of \textsl{XMM-Newton}\footnote{The current calibration documentation of \textsl{XMM-Newton} EPIC cameras can be downloaded from http://xmm2.esac.esa.int/docs/documents/CAL-TN-0018.pdf}, and the final results of the fitting are consistent with those presented in Table 2. Therefore, we decided not to apply this energy shift to the \textsl{XMM-Newton} PN data in the final fitting of all models.

\begin{figure}
\epsscale{1.15}
\plotone{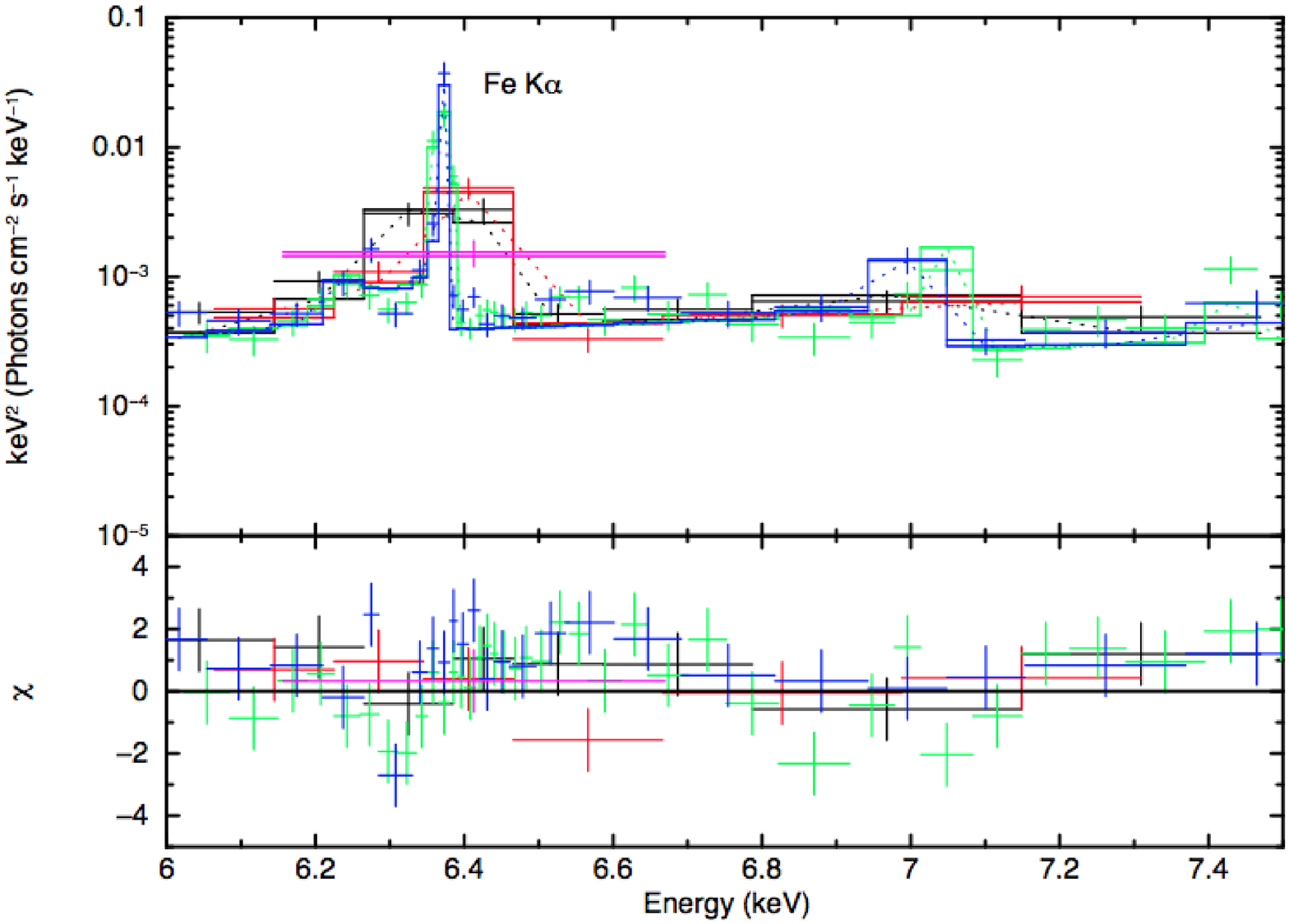}
\caption{A zoom in of the spectral fit for Model T between 6--7.5 keV showing the asymmetric residuals around the iron line emission; the x-axis has been de-redshifted to show the rest-frame energy. The color scheme of the data and the model are the same as Figure 2. The residuals are dominated by the \textsl{XMM-Newton} data at $\approx$6.3--6.8 keV, with the model overestimating the emission at $\approx$6.3 keV, and underpredicting the data at $\approx$6.4--6.8 keV. These residuals are diminished when applying an $\sim +11$ eV offset in the \textsl{XMM-Newton} PN data.}
\end{figure}


\subsection{Model M}

We next fitted the AGN spectrum using the MYT{\sc{orus}} model by \citet{MY09} (Model M). This model simulates obscuration toward an AGN using a toroidal absorber geometry (circular cross-section) with a fixed opening angle $\theta_{\rm{tor}}$ $=$ 60$^{\circ}$. The advantage of this model over Model T is that it allows the separation of the direct ({\sc{mytz}}), scattered ({\sc{myts}}) and line emission ({\sc{mytl}}) components, allowing more freedom to explore complex geometries in the modeling. On the other hand, the line-of-sight $N_{\rm{H}}$(los) is tied to the inclination angle and the equatorial $N_{\rm{H}}$(eq), which is available only up to an absorbing column density of $N_{\rm{H}}$(eq) $=$ 10$^{25}$ cm$^{-2}$. The {\sc{mytl}} component self-consistently includes neutral Fe K$\alpha$ and Fe K$\beta$ fluorescence lines, and the associated Compton shoulders. The model is defined between 0.5--500 keV, but we noticed strong residuals in our fit near the lower energy threshold, and therefore restricted our fits to above 0.6 keV for this model.\footnote{Since the fit for Model M was restricted to $E$ $\geq$ 0.6 keV, a Gaussian component that was added to model the emission line at $E$ $\approx$ 0.58 keV (M13) in Model P and T was excluded for this model.}

We fitted the AGN spectrum using the simplest version of the MYT{\sc{orus}} model, which couples all the parameters of the scattered and fluorescence line components to the direct continuum component. We added an extra Gaussian component to account for the Ni K$\alpha$ line which is not included in the model (parameters were fixed to values obtained in M13). The relative normalizations of {\sc{myts}} ($A_{S}$) and {\sc{mytl}} ($A_{L}$) with respect to {\sc{mytz}} ($A_{Z}$) were set to 1. Model M also fits the data well with $\chi^{2}$/d.o.f $=$ 569/471. The best-fitting global column density inferred is $N_{\rm{H}}$(eq) $=$ 10.0$^{+u}_{-0.5}$ $\times$ 10$^{24}$ cm$^{-2}$, which is at the upper limit of the MYT{\sc{orus}} model. Reassuringly, the best-fitting inclination angle is $\theta_{\rm{inc}}$ $=$ 65.9$^{+2.33}_{-1.87}$ degrees, consistent with that determined by \citet{Fischer13}. However, we note that the MYT{\sc{orus}} model does not allow for complete freedom in the fitting of the inclination angles. Due to the assumption made for $\theta_{\rm{tor}}$, the model treats a torus with $\theta_{\rm{inc}}$ $<$ 60$^{\circ}$ as unobscured, and as a result, the best-fit $\theta_{\rm{inc}}$ measured by the model for obscured AGN are usually in the range of $\approx$60--70$^{\circ}$. The \emph{N}$_{\rm H}$(\rm los) measured from this model is well within the CT regime, with an unconstrained upper limit; i.e., \emph{N}$_{\rm H}$(\rm los) $=$ 5.8$^{+u}_{-1.2}$ $\times$10$^{24}$ cm$^{-2}$. Similar to Model T, the residuals of the fit are dominated by the soft energy emission, and the emission around the iron line, which can be improved if an energy offset is allowed in the \textsl{XMM-Newton} PN data (see Section 3.2). All parameters obtained by Model M, including the cross-calibration normalization constants and intrinsic luminosities of the AGN, agree very well with Model T (see Table 2).

\subsection{The {\sc{sphere}} Model}

In addition to the {\sc{torus}} model described in Section 3.2, \citet{BN11} also present a model in which the source is fully covered by a spherical geometry ($\theta_{\rm{tor}} =$ 0$^{\circ}$) with variable elemental and iron abundances with respect to hydrogen. This model is referred to as the {\sc{sphere}} model. We tried to fit the spectra using this model. However, the fit was very poor ($\chi^{2}$/d.o.f $=$ 1169/501) and difficult to constrain. Leaving the metal abundances free to vary, which is allowed in this model, did improve the fit slightly, but the fit was still poor ($\chi^{2}$/d.o.f $=$ 848/499), and returned very low values of $N_{\rm{H}}$ and $\Gamma$, and a high Fe abundance ($\sim$ 7). Therefore, we will not discuss this model any further.   


\section{NGC 5643 X-1} 


\begin{figure*}
\epsscale{0.55}
\plotone{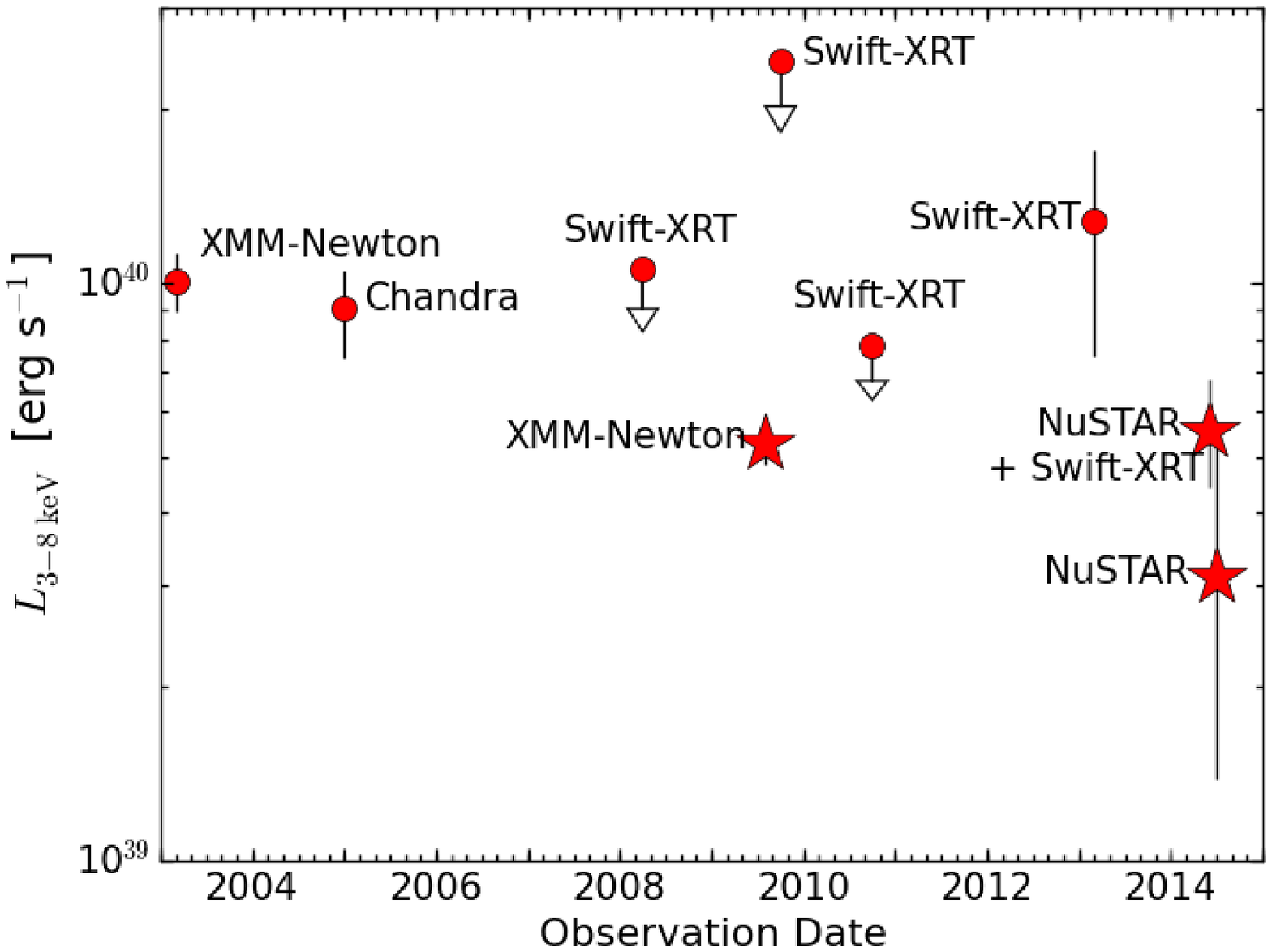}
\epsscale{0.53}
\plotone{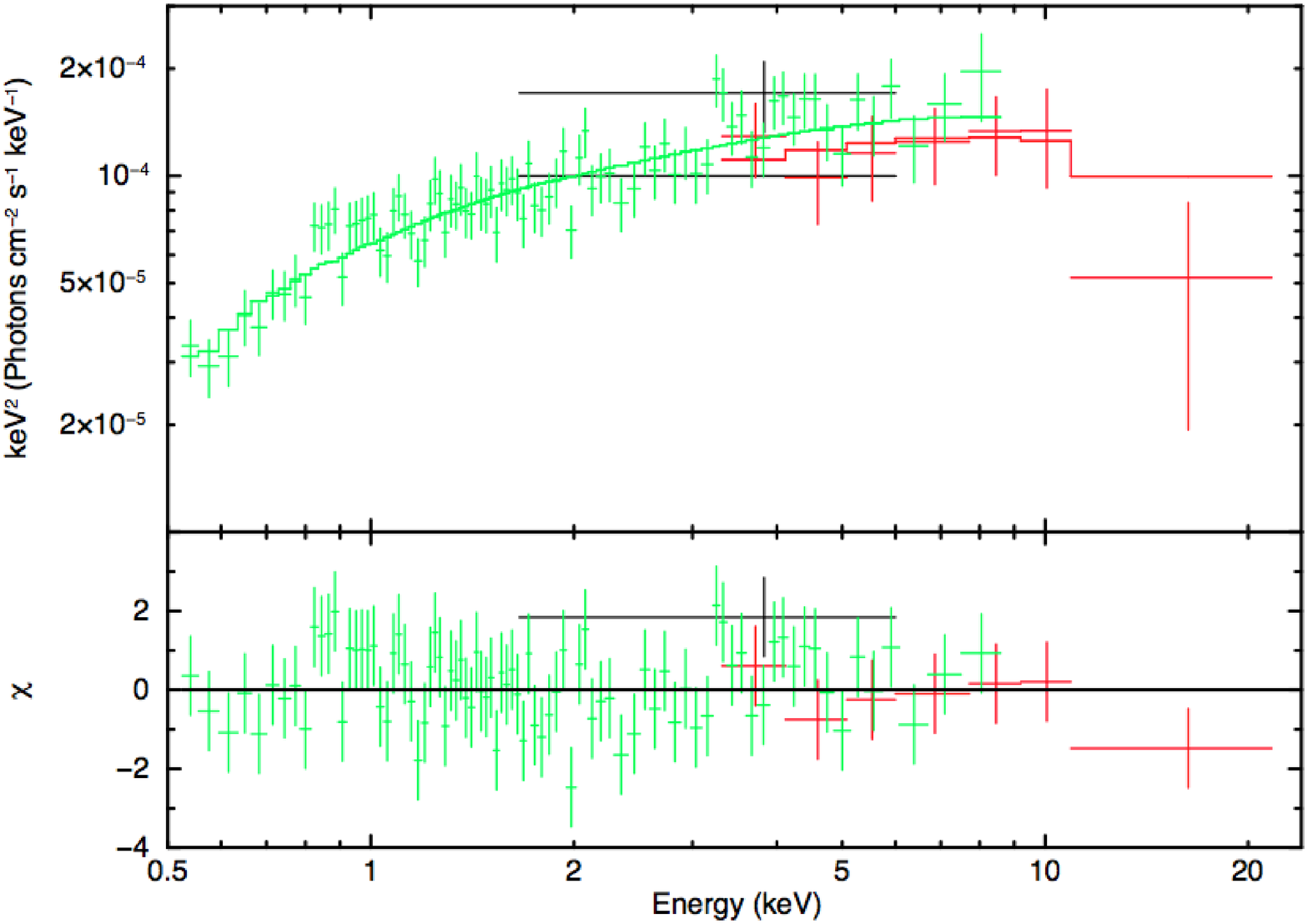}
\caption{Left: X-ray luminosity variations of NGC 5643 X--1 in the 3--8 keV band observed between 2003--2014 by \textsl{XMM-Newton}, \textsl{Chandra}, \textsl{Swift}-XRT and \textsl{NuSTAR}. Observations marked with stars are those that we used to fit the broadband spectrum of the source. Right: Best-fit {\sc{diskpbb}} model to the combined \textsl{Swift}-XRT+\textsl{XMM}+\textsl{NuSTAR} spectra of NGC 5643 X--1. The top panel shows the unfolded model in $EF_{E}$ units, while the bottom panel shows the fit residuals in terms of sigma with error bars of size one. The data have been rebinned to a minimum of 3-sigma significance with a maximum of 50 bins for visual clarity. Color scheme: red (\textsl{NuSTAR}), black (\textsl{Swift}-XRT), green (\textsl{XMM-Newton} MOS).}
\end{figure*}

\begin{table*}
\begin{center}
\caption{X-ray spectral fitting results for NGC 5643 X--1.}
\begin{tabular}{lcccc}
\tableline
& & &\\
 Parameter   & Unit       & \textsl{Swift}-XRT+\textsl{NuSTAR}  & \textsl{Swift}-XRT+\textsl{XMM}+\textsl{NuSTAR}\\
& & &\\
\tableline
  Model = {\sc{tbabs}} $\times$ {\sc{zpowerlaw}}       &                  &  &      \\
\tableline
  $N_{\rm{H}}$ & 10$^{21}$ cm$^{-2}$ &  $<$ 23.2 & $<$ 0.46    \\
  $\Gamma$ &                                    &   2.35$_{-0.33}^{+0.44}$  & 1.70$_{-0.08}^{+0.09}$    \\
   $L_{0.5-24}$   & 10$^{40}$ erg s$^{-1}$    & 1.18$_{-0.31}^{+0.16}$ &  1.22$_{-0.03}^{+0.07}$            \\
   $C_{\rm{NuSTAR}}^{\rm{XMM}}$ &  & -          &  1.31$_{-0.23}^{+0.31}$                \\         
 $\chi_{\rm{r}}^{2}$/d.o.f.   &       & 1.32/7 & 1.08/92     \\
\tableline
   Model = {\sc{tbabs}} $\times$ {\sc{cutoffpowerlaw}}                                       &  &   &    \\
\tableline
  $N_{\rm{H}}$ & 10$^{21}$ cm$^{-2}$  & $<$ 18.5   & $<$ 0.22  \\
  $\Gamma$ &                                 &    1.70$_{-1.64}^{+0.98}$ & 1.52$_{-0.11}^{+0.13}$      \\
  $E_{\rm{cut}}$ & keV                    &   10.7$_{-7.6}^{+u}$  & 16.3$_{-9.6}^{+35.9}$     \\
   $L_{0.5-24}$   & 10$^{40}$ erg s$^{-1}$         &    1.08$_{-0.11}^{+0.22}$           &  1.18$\pm$0.07            \\
   $C_{\rm{NuSTAR}}^{\rm{XMM}}$ &  &         -       &  1.12$_{-0.20}^{+0.36}$                \\   
$\chi_{\rm{r}}^{2}$/d.o.f.   &        &  1.41/6 & 1.03/91      \\
\tableline
   Model = {\sc{tbabs}} $\times$ {\sc{diskpbb}}                                &   & &      \\
\tableline
  $N_{\rm{H}}$ & 10$^{21}$ cm$^{-2}$ & $<$ 16.8 & ...     \\
  $T_{\rm{in}}$ & keV  &    4.02$_{-1.47}^{+4.81}$  & 4.98$_{-0.20}^{+0.17}$     \\
  $p$ &  &                  $<$ 0.74 & 0.55$\pm$0.01  \\
   $L_{0.5-24}$   & 10$^{40}$ erg s$^{-1}$         & 1.01$_{-0.20}^{+0.36}$    &  1.07$_{-0.08}^{+0.06}$    \\
   $C_{\rm{NuSTAR}}^{\rm{XMM}}$ &  &    -            &      1.19$_{-0.23}^{+0.24}$            \\   
$\chi_{\rm{r}}^{2}$/d.o.f.   &        &  1.28/6   & 0.99/91  \\
\tableline
\end{tabular}
\tablecomments{$^{u}$unconstrained.}
\end{center}
\end{table*}



In addition to the AGN, we also analyzed the \textsl{NuSTAR} data for the ULX candidate, NGC 5643 X--1. ULXs are off-nuclear point sources with X-ray luminosities exceeding the Eddington limit for the typical $\sim$10$M_{\odot}$ stellar-remnant black holes observed in Galactic black hole binaries (e.g. \citealp{Orosz03}); i.e., $L_{\rm{X}}$ $\gtrsim$ 10$^{39}$ erg s$^{-1}$, potentially implying exotic super-Eddington accretion (e.g. \citealp{Poutanen07}). While the majority of ULXs have luminosities of the order of 10$^{39}$ erg s$^{-1}$ (\citealp{Walton11a}; \citealp{Swartz11}), a small subset of the population has been observed to have $L_{\rm{X}}$ $>$ 10$^{40}$ erg s$^{-1}$ (\citealp{Walton11b}; \citealp{Jonker12}; \citealp{Sutton12}). Given their luminosities, it has been suggested that these more luminous sources might be good candidates for hosting intermediate mass black holes (10$^{2}$ $\lesssim$ $M_{\odot}$ $\lesssim$ 10$^{5}$; e.g. \citealp{Miller04}; \citealp{Strohmayer09}) accreting at sub-Eddington rates. With a luminosity of \emph{L}$_{0.5-10}$ $=$ 2.6 $\times$ 10$^{40}$ erg s$^{-1}$ as measured by G04, NGC 5643 X--1 would therefore be a member of this latter population.

We extracted the \textsl{NuSTAR} spectra of NGC 5643 X--1 by defining a smaller circular region of 20$\arcsec$-radius around the source using the AGN as a reference point for its relative position. For the background region, we used an annular segment centered on the AGN (excluding a region of $\sim$50$\arcsec$-radius around NGC 5643 X--1 which accounts for the $\sim$70$\%$ encircled energy fraction), and with the same radial distance (relative to the AGN position) and width as the source region. This background region was designed to account for contamination by the AGN. We then determined the significance of the \textsl{NuSTAR} detection by calculating a no-source probability assuming binomial statistics (\textsl{P} $>$ 1$\%$ for non-detection; i.e. $\lesssim$ 2.6$\sigma$, following other \textsl{NuSTAR} studies of faint sources -- e.g. \citealp{Luo13}; \citealp{Lansbury14}; \citealp{Stern14}). While the source is significantly detected in both of the \textsl{NuSTAR} observations in the 3--8 keV band, it is only significantly detected in the second observation in the 8--24 keV band (\textsl{P} $=$ 2.21 $\times$ 10$^{-6}$ in the combined FPMA+B image). 

We next combined the data from the two FPMs for each epoch and binned the spectra to a minimum of 5 counts per bin due to low counts. We then fitted the spectra from each epoch using a simple power-law model, absorbed by the Galactic column. The spectral fitting parameters were calculated using C-statistics, appropriate for low count statistics \citep{Cash79}. Although the source is not formally detected in the 8--24 keV band in the first epoch, we modeled the spectrum including the data point in this band to provide a better constraint to the fit. We found that the spectra of the source during the first and second epoch are consistent with each other, $\Gamma$ $=$ 2.7$_{-0.5}^{+0.6}$ and $\Gamma$ $=$ 1.7$_{-0.5}^{+0.6}$, \emph{f}$_{3-24}$ $=$ 3.6$_{-0.8}^{+0.9}$ $\times$ 10$^{-13}$ erg s$^{-1}$ cm$^{-2}$ and 3.8$_{-1.0}^{+1.4}$ $\times$ 10$^{-13}$ erg s$^{-1}$ cm$^{-2}$, respectively, suggesting that the source has not varied significantly between the two observations. The lack of a detection at 8--24 keV in the first epoch may be due to the location of NGC 5643 X--1 relative to the \textsl{NuSTAR} detector gap (Figure 1), which reduced the overall sensitivity of the data.

Motivated by the lack of significant flux variability, we therefore combined the \textsl{NuSTAR} spectra of NGC 5643 X--1 from the two epochs and fitted the total spectrum with the \textsl{Swift}-XRT data that were taken simultaneously with the first epoch. We reduced the \textsl{Swift}-XRT data as detailed in Section 2.4, and extracted the spectrum using a 20$\arcsec$-radius circular source region to match the \textsl{NuSTAR} extraction region. The background photons were selected from a source-free 100$\arcsec$-radius circular region close to the source. We grouped the data to a minimum of 40 and 20 counts per bin for \textsl{NuSTAR} and \textsl{Swift}-XRT, respectively, allowing the use of $\chi^{2}$ statistics.

We first fitted the broadband spectrum of NGC 5643 X--1 over 0.5--24 keV using a simple power-law model absorbed by a column density intrinsic to the host galaxy ({\sc{tbabs}}) in addition to the Galactic absorption. We fixed the cross-normalization constants between \textsl{NuSTAR} and \textsl{Swift}-XRT to 1. The fit is acceptable with $\chi^{2}$/d.o.f $=$ 9.21/7. The spectrum seems to drop off at E $\approx$ 11 keV, which could be an indication of a spectral cut-off at around 10 keV, as found for other \textsl{NuSTAR}-observed ULXs (e.g. \citealp{Walton13}; \citealp{Bachetti13}; \citealp{Walton14}; \citealp{Rana15}). If shown to be statistically significant, this would be the first time that such a cut-off is observed for NGC 5643 X--1. We therefore proceeded to fit the spectrum using an absorbed cut-off power-law model. The fit is also acceptable ($\chi^{2}$/d.o.f $=$ 8.46/6), and returned a cut-off energy of $E_{\rm{cut}}$ $=$ 10.7$_{-7.6}^{+u}$ keV. We also fitted the spectra using a multi-color blackbody accretion disk model with a variable temperature disk profile, {\sc{diskpbb}} \citep{Mineshige94}, which is commonly used to model ULX spectra. This model provides a marginal improvement to the fit over the simple and cut-off power-law model ($\chi^{2}$/d.o.f $=$ 7.67/6). The measured temperature profile, $p$, is less than that expected for a thin disk ($p$ $<$ 0.75; \citealp{Shakura73}), consistent with an accretion disk in which advection of radiation is important, as expected at very high accretion rates where radiation pressure dominates and modifies the structure of the disk (e.g. \citealp{Abramowicz88}). This would be consistent with NGC 5643 X--1 exhibiting high-Eddington rate accretion onto a stellar remnant black hole. The luminosity inferred by this model is $L_{0.5-24}$ $=$ 1.01$_{-0.20}^{+0.36}$ $\times$ 10$^{40}$ erg s$^{-1}$, assuming that NGC 5643 X--1 is at the distance of NGC 5643.


To obtain a better constraint to the broadband spectral fit of NGC 5643 X--1, we fitted the \textsl{Swift}-XRT and \textsl{NuSTAR} spectra simultaneously with the 2009 \textsl{XMM-Newton} spectrum of the source, in which M13 measured a luminosity of $L_{2-10}$ $=$ (8.11 $\pm$ 0.23) $\times$ 10$^{39}$ erg s$^{-1}$, consistent with that inferred by our \textsl{Swift}-XRT+\textsl{NuSTAR} best-fit {\sc{diskpbb}} model $L_{2-10}$ $=$ 7.53$_{-1.39}^{+1.21}$ $\times$ 10$^{39}$ erg s$^{-1}$. Data were reduced as described in Section 2.2. Because NGC 5643 X--1 fell on the detector gap in the PN camera, we only extracted the spectra of the source from the two MOS cameras, and co-added the spectra together. We extracted the spectra using a 20$\arcsec$-radius circular source region, and a 45$\arcsec$-radius circular background region. We left the cross-normalization between the \textsl{XMM-Newton} spectrum relative to \textsl{NuSTAR}, $C_{\rm{NuSTAR}}^{\rm{XMM}}$, free to vary. 

We found that both the cut-off power-law and {\sc{diskpbb}} models provide better fits to the \textsl{Swift}-XRT+\textsl{XMM}+\textsl{NuSTAR} data than a simple power-law model ($\chi^{2}$/d.o.f $=$ 93.6/91, 90.2/91 and 99.0/92, respectively). The F-test null hypothesis probability between the simple power-law and cut-off power-law model is 0.023. This low value is strongly suggestive of the need for the high energy cut-off, though deeper observation will be required to validate this. $C_{\rm{NuSTAR}}^{\rm{XMM}}$ measured by the two models are, within the uncertainties, consistent with 1, providing further evidence that the spectra are in good agreement with each other. The cut-off power-law model measured a cut-off energy of $E_{\rm{cut}}$ $=$ 16.3$_{-9.6}^{+35.9}$ keV, and the {\sc{diskpbb}} model inferred a temperature disk profile, $p$ $=$ 0.55$\pm$0.01 and an inner disk temperature, $T_{\rm{in}}$ $=$ 4.98$_{-1.80}^{+2.46}$ keV. This temperature is consistent with that measured for some ULXs when fitting their whole broadband spectrum with just a single disk component (e.g. \citealp{Walton14}; \citealp{Rana15}). The best-fitting parameters for all models are presented in Table 3, and the broadband \textsl{Swift}-XRT+\textsl{XMM}+\textsl{NuSTAR} spectrum of NGC 5643 X--1 fitted by the best-fit model, namely {\sc{diskpbb}}, is shown in Figure 4.

Strong X-ray luminosity variations are often observed in ULXs. Motivated by this, we also reduced the 2003 \textsl{XMM-Newton} data (ObsID 0140950101; UT 2003-02-08), as well as the \textsl{Chandra} data, and fitted the 3--8 keV spectra of NGC 5643 X--1 using a simple power-law model absorbed by the Galactic column, to infer its 3--8 keV luminosities at these epochs. In addition to these observations, the galaxy has also been observed by \textsl{Swift}-XRT between 2008--2013 on $\sim$30 occasions (mostly $\sim$1 ks exposure time), with most observations close to each other in 2013 targetted on an ongoing supernova in the galaxy. For simplicity, for each year in which there are multiple \textsl{Swift}-XRT observations of the source, we only analyzed the data with the highest nominal exposure time to provide the best estimate of NGC 5643 X--1 luminosity in that particular year (ObsID 00037275001, 00037275002, 00037275004 and 00032724009; UT 2008-03-16, 2009-09-20, 2010-10-01 and 2013-02-26, respectively). Among these observations, NGC 5643 X--1 is only significantly detected by \textsl{Swift}-XRT in the 2013 observation. Therefore for this observation, we extracted the data as described earlier, and fitted the 3--8 keV spectrum using a simple power-law model absorbed by the Galactic column. For the other observations where NGC 5643 X--1 is not detected, we provided the upper limit luminosities at 90$\%$ confidence, estimated using aperture photometry assuming $\Gamma$ $=$ 2. 

We compare the 3--8 keV luminosities between all the observations described above, as well as the \textsl{NuSTAR} observations at each epoch in Figure 4 to show the long term luminosity variability of NGC 5643 X--1. Note that the photon indices were left free to vary in all spectral fittings. In general, the plot shows that NGC 5643 X--1 can vary by a factor of $\sim$2--3 (up to $\sim$5) between epochs. This is broadly similar to the level of variability observed in other ULXs (e.g. \citealp{Kong10}; \citealp{Sutton13}; \citealp{Walton13}). These results, combined with the parameters inferred by the {\sc{diskpbb}} model and evidence of an energy cut-off at E $\sim$ 10 keV, as well as the lack of a counterpart at other wavelengths (G04), further support the ULX classification of NGC 5643 X--1.



\section{Discussion}

In Section 3, we first investigated \textsl{NuSTAR} observations of the CTAGN candidate in NGC 5643. We combined our new data with archival \textsl{Chandra}, \textsl{XMM-Newton} and \textsl{Swift}-BAT data, and performed a broadband ($\sim$0.5--100 keV) spectral analysis of the AGN. On the basis of three different reflection/obscuration models, we found that the AGN is consistent with being CT with a column density of $N_{\rm{H}}$(los) $\gtrsim$ 5 $\times$ 10$^{24}$ cm$^{-2}$. The absorption-corrected 2--10 keV luminosity obtained from the models range between \emph{L}$_{2-10}$ $=$ (0.8--1.7) $\times$ 10$^{42}$ erg s$^{-1}$, although we note that this could be a factor of a few lower/higher due to the fact that the spectra are dominated by reflection with negligible contribution from the direct emission, causing large uncertainties in the measurements of the intrinsic emission from the AGN. The intrinsic luminosity inferred is at the lower end of the luminosity range of the local bona-fide CTAGN population (see Figure 4 in \citealp{Gandhi14}). Despite the presence of a nearby ULX with a 2--10 keV luminosity comparable to the observed luminosity of the AGN itself (G04), the spectrum of the ULX drops off at $E$ $\sim$ 10 keV, and the overall emission above this energy is dominated by the AGN. 

This paper provides the first intrinsic X-ray luminosity measurements for the AGN in NGC 5643 using Compton-thick solutions. To provide a basic test of the intrinsic luminosity estimate from our X-ray spectral fitting, we can compare the luminosity obtained to predictions from multiwavelength indicators. The obscuring torus absorbs disk radiation from the nucleus, but then re-emits it in the MIR. We can therefore use the intrinsic MIR:X-ray luminosity relationship found by \citet{Gandhi09} based upon high angular resolution MIR 12$\mu$m observations to estimate the intrinsic X-ray luminosity of the AGN. The advantage of such observations is that because of their high angular resolution ($\sim$0.4$\arcsec$), they provide MIR fluxes that are intrinsic to the nucleus, minimizing contamination from the host galaxy. The 12$\mu$m nuclear luminosity for NGC 5643, measured from VLT VISIR and Gemini T-ReCS observations, is \emph{L}$_{12\mu m}$ $=$ (1.5$\pm$0.4) $\times$ 10$^{42}$ erg s$^{-1}$ \citep{Asmus14}. Using the \citet{Gandhi09} relationship, this corresponds to an X-ray luminosity of \emph{L}$_{2-10}$ $=$ (0.6--2.5) $\times$ 10$^{42}$ erg s$^{-1}$, which agrees well with the 2--10 keV intrinsic luminosity inferred from our broadband X-ray spectral modeling.\footnote{The given luminosity range accounts for the mean scatter of the correlation.} 

\begin{figure}
\epsscale{1.18}
\plotone{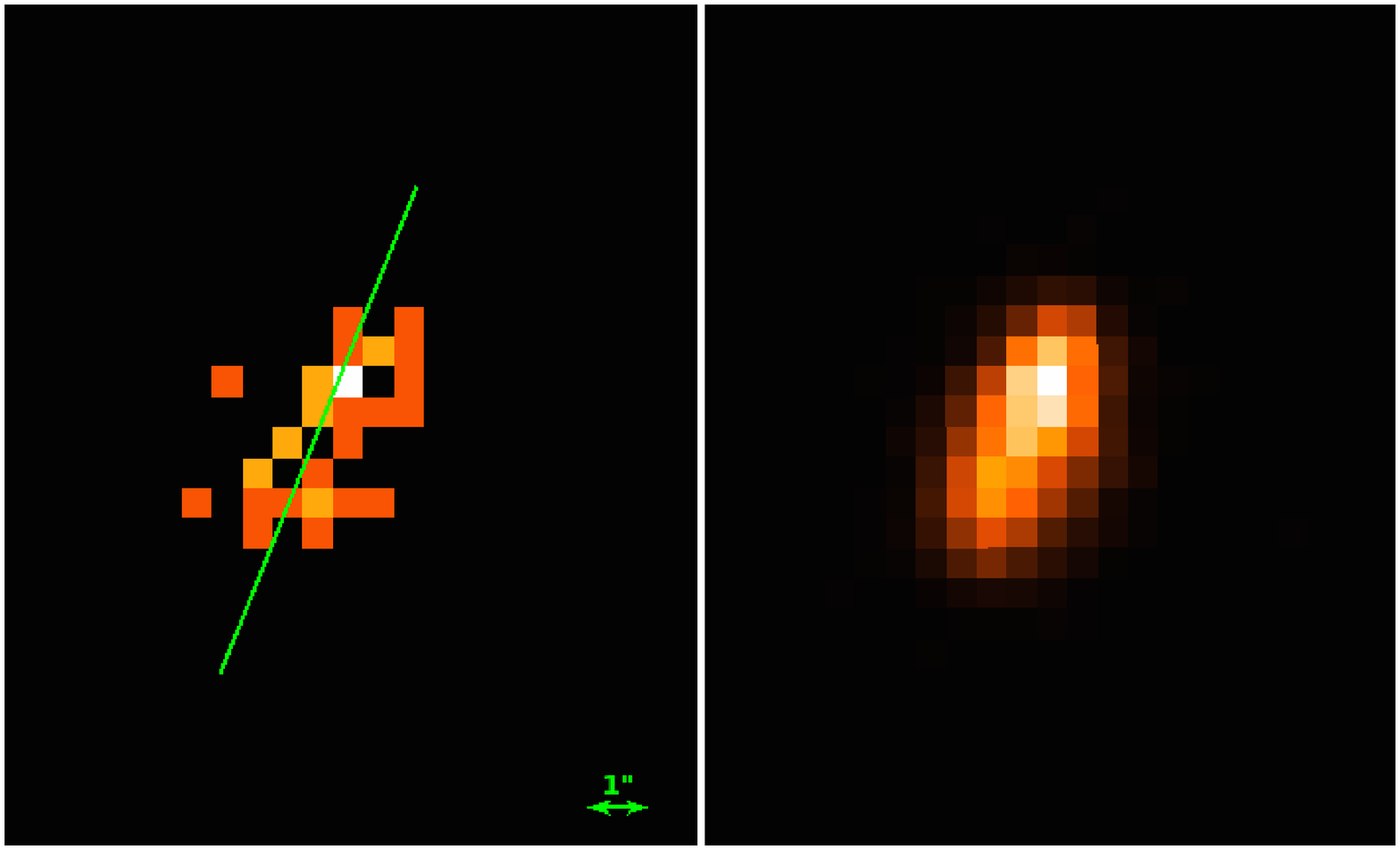}
\epsscale{1.2}
\plotone{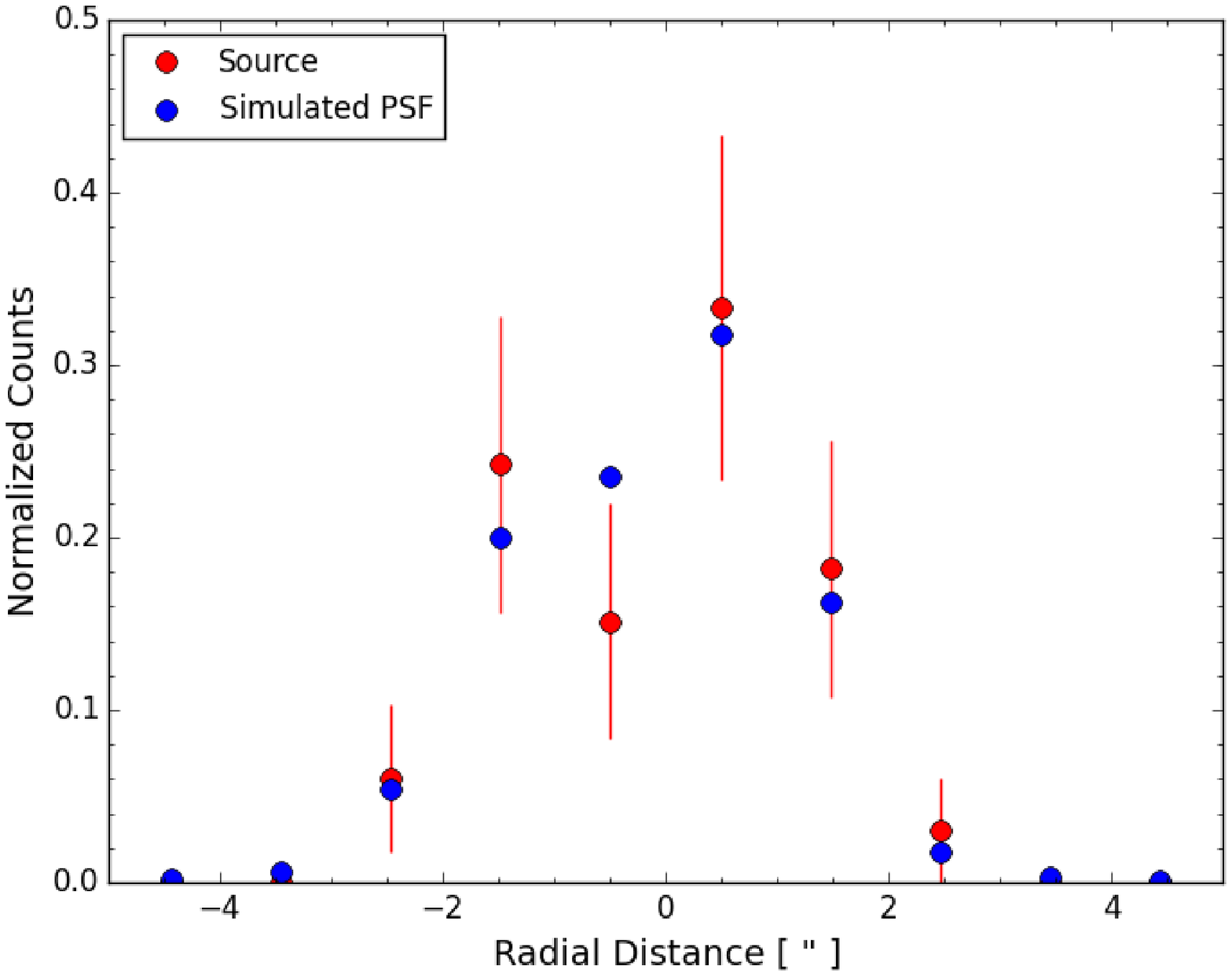}
\caption{Top: \textsl{Chandra} 6--7.5 keV images of the AGN (left) and the simulated PSF (right). The green line shows the direction of the semi-major axis of the simulated PSF that was used to produce the 1D radial profiles for both the source and the simulated PSF. Bottom: The 1D radial profiles of the source (red) and the simulated PSF (blue) normalized to their respective total counts. The error bars indicate the 1-$\sigma$ poisson errors for the data.}
\end{figure}

We also compared our intrinsic 2--10 keV luminosity with that predicted from the optical [O{\sc{iii}}]$\lambda$5007\AA \ emission line luminosity. [O{\sc{iii}}]$\lambda$5007\AA \ in AGN are mostly produced in the NLR by photoionizing radiation from the central source. Although this line is not affected by the torus obscuration, it can still suffer from obscuration by the host galaxy, which can be corrected for using the Balmer decrement. We obtained our corrected [O{\sc{iii}}]$\lambda$5007\AA \ luminosity from \citet{Bassani99}, $\mathrm{\emph{L}_{[O{\sc{III}}]}}= ($1.6$\pm$0.2) $\times$ 10$^{41}$ erg s$^{-1}$. Using the \emph{L}$_{2-10}$:$\mathrm{\emph{L}_{[O{\sc{III}}]}}$ relationship from \citet{Panessa06}, we infer an X-ray luminosity of \emph{L}$_{2-10}$ $=$ (2.1--33.9) $\times$ 10$^{42}$ erg s$^{-1}$.$^{9}$ The lower end of this luminosity range is consistent with the luminosity measured from our X-ray spectral fitting. This is supported by the MIR [O{\sc{iv}}]$\lambda$25.89$\mu$m emission line, $\mathrm{\emph{L}_{[O{\sc{IV}}]}} =  ($2.7$\pm$0.1) $\times$ 10$^{40}$ erg s$^{-1}$ \citep{Goulding09}. Using \emph{L}$_{2-10}$:$\mathrm{\emph{L}_{[O{\sc{IV}}]}}$ relationship defined by \citet{Goulding10}\footnote{We assume a typical AGN bolometric correction \emph{L}$_{\rm{bol}}$/\emph{L}$_{2-10}$ $\approx$ 20 to calculate the bolometric luminosity and hence the mass accretion rate of the AGN (e.g. \citealp{Elvis94}; \citealp{Vasudevan10}).}, we determine an X-ray luminosity of \emph{L}$_{2-10}$ $=$ (0.9--5.5) $\times$ 10$^{42}$ erg s$^{-1}$.$^{9}$ This luminosity agrees very well with that measured by our best-fit models.

With our updated intrinsic X-ray luminosity, we can estimate the Eddington ratio of the AGN. Using \emph{M}$_{\rm{BH}}$ estimated from the \emph{M}$_{\rm{BH}}$--$\sigma_{*}$ relation, \emph{M}$_{\rm{BH}}$ $=$ 10$^{6.4}$ M$_{\odot}$ (\citealp{Goulding10}; see also footnote 3), combined with the range of intrinsic X-ray power that we measured from the best-fit models, we infer that the AGN is accreting matter at about 5--10$\%$ of the Eddington rate.$^{11}$ However, we note that this value is subject to large errors (factor of a few) due to the highly uncertain quantities involved in the calculation \citep{BrandtAlexander15}.

\begin{figure}
\epsscale{1.17}
\plotone{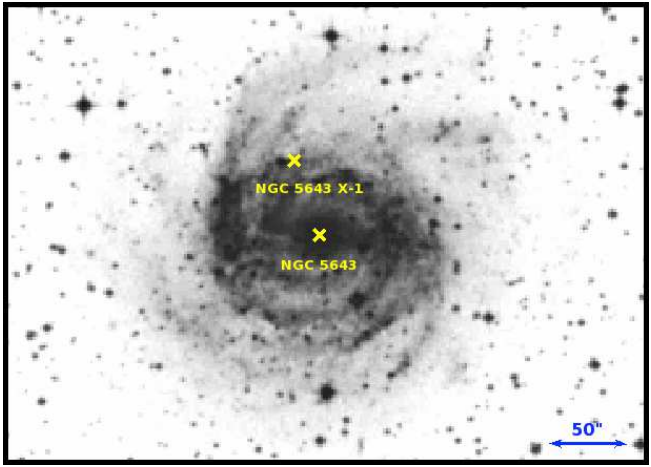}
\caption{Optical image of NGC 5643 retrieved from the ESO Digital Sky Survey image server. \textsl{Chandra} position of the AGN (NGC 5643) and the ULX candidate (NGC 5643 X--1) are both labeled and marked with an ``$\times$". North is up and East is to the left in the image.} 
\end{figure}

In the past few years, much evidence for the presence of distant cold reflecting regions have been found in nearby CTAGN (e.g. \citealp{Marinucci13}; \citealp{Arevalo14}; \citealp{Bauer14}). To accurately characterize the broadband spectra of these AGN, multiple reflector components are required to model each of these reflecting regions. We investigated whether this could be the case for NGC 5643 by looking at the \textsl{Chandra} image of the AGN in the 6--7.5 keV band ($\sim$30 counts). This is shown in the top left panel of Figure 5. At first glance, the image appears to show that the iron emission region is extended well beyond the parsec-scale torus, suggesting the presence of distant cold reflecting regions. However, this appearance could also be due to an elongated PSF as a result of the position of the target on the detector, as well as the energy range used to produce the source image. We therefore used the MARX simulator\footnote{The MARX simulation documentation is available at http://space.mit.edu/ASC/MARX/.} (v4.4) to produce the expected PSF with a much higher count than the source counts ($\sim$ 100 000 counts) at the position of the AGN, and at the mean energy range that we are interested in ($E$ $=$ 6.75 keV). This is shown in the top right panel of Figure 5. We then plotted 1D radial profiles of the target and the simulated PSF along the semi-major axis of the simulated PSF determined using the {\sc{wavdetect}} tool within {\sc{ciao}}, and performed a Kolmogorov-Smirnov (KS) test between the two distributions (bottom panel of Figure 5). We found that the KS test probability is $P_{\rm{KS}}$ $=$ 0.983, indicating that the iron emission region of the target is consistent with the simulated PSF, and therefore is not extended (see similar finding for NGC 3393 in \citealp{Koss15}). A deeper (scheduled) \textsl{Chandra} observation could help to confirm this.


The CTAGN in NGC 5643 is located in a face-on grand design spiral galaxy with an undisturbed disk (see Figure 6). The emission lines from the AGN are not obscured/extinguished by the host galaxy material \citep{Goulding09}, indicating that any phenomenon that might contribute to the CT obscuration in this source is therefore likely to occur very close to the nucleus. Indeed, there have been indications that the emission at the west side of the nucleus is obscured by a warped disk or the presence of a dust lane (e.g. \citealp{Simpson97}, \citealp{Fischer13}, \citealp{Davies14}). The torus itself could be the inner structure of this disk. Because of the more direct and ``cleaner'' view of the central engine in NGC 5643 as compared to some other very nearby bona-fide CTAGN such as NGC 1068, NGC 4945 and Circinus, this makes NGC 5643 a good candidate for more detailed studies of the CT obscurer in the nuclear region.


\section{Summary}

We observed the AGN and ULX candidate in NGC 5643 using \textsl{NuSTAR} on two occasions conducted at about a month separation. A summary of our main results is as follows:

\begin{enumerate}
 \item Using the combined data from \textsl{NuSTAR}, \textsl{Chandra}, \textsl{XMM-Newton} and \textsl{Swift}-BAT, we have extended the broadband spectral analysis of the CTAGN candidate in NGC 5643 to high energies ($\sim$0.5--100 keV). Using physically motivated toroidal obscuration models, we showed that the source is indeed CT with a column density of $N_{\rm{H}}$(los) $\gtrsim$ 5 $\times$ 10$^{24}$ cm$^{-2}$. 
 \item The absorption-corrected 2--10 keV luminosity measured by these models is \emph{L}$_{2-10,\rm{int}} =$ (0.8--1.7) $\times$ 10$^{42}$ erg s$^{-1}$, consistent with that predicted from multiwavelength intrinsic luminosity indicators. The luminosity inferred is at the lower end of the luminosity range of the local CTAGN population. 
 \item The \textsl{NuSTAR} spectra of the off-nuclear source, NGC 5643 X-1, shows evidence for a spectral cut-off at $E$ $\sim$ 10 keV, similar to that observed in other ULXs observed by \textsl{NuSTAR}. Combining this information with the evidence for X-ray luminosity variations observed between different observations, along with the absence of unambiguous counterparts at other wavelengths, we concluded that the source is consistent with being a ULX. Future simultaneous low and high energy X-ray observations of this field are needed in order to provide higher quality data to confirm the spectral cut-off that we observed, and to better characterize the broadband spectrum of the source. 
\end{enumerate}


\section*{Acknowledgments}
The authors thank the anonymous referee for useful comments which have helped to improve the paper. We thank Chris Done for some discussions on the residuals around the iron line complex. A.A thanks Wasutep Luangtip for useful discussion on NGC 5643 X--1, and help with the MARX simulation. We also thank Neil Gehrels and the \textsl{Swift} team for the simultaneous \textsl{Swift}-XRT observation. We acknowledge financial support from Majlis Amanah Rakyat (MARA) Malaysia (A.A), the Science and Technology Facilities Council (STFC) grant ST/J003697/1 (P.G), ST/I0015731/1 (D.M.A and A.D.M), ST/K501979/1 (G.B.L), the Leverhulme Trust (D.M.A), NASA Headquarters under the NASA Earth and Space Science Fellowship Program grant NNX14AQ07H (M.B), CONICYT-Chile grants Basal-CATA PFB-06/2007 (F.E.B), FONDECYT 1141218 (F.E.B), and ``EMBIGGEN" Anillo ACT1101 (F.E.B); the Ministry of Economy, Development, and Tourism's Millennium Science Initiative through grant IC120009, awarded to The Millennium Institute of Astrophysics, MAS (F.E.B). 

\textsl{NuSTAR} is a project led by the California Institute of Technology (Caltech), managed by the Jet Propulsion Laboratory (JPL), and funded by the National Aeronautics and Space Administration (NASA). We thank the \textsl{NuSTAR} Operations, Software and Calibrations teams for support with these observations. This research has made use of the \textsl{NuSTAR} Data Analysis Software ({\sc{nustardas}}) jointly developed by the ASI Science Data Center (ASDC, Italy) and the California Institute of Technology (USA), and the XRT Data Analysis Software ({\sc{xrt-das}}). This research also made use of the data obtained through the High Energy Astrophysics Science Archive Research Center (HEASARC) Online Service, provided by the NASA/Goddard Space Flight Center, and the NASA/IPAC extragalactic Database (NED) operated by JPL, Caltech under contract with NASA. 

\textsl{Facilities}: \textsl{NuSTAR}, \textsl{XMM-Newton}, \textsl{Chandra}, \textsl{Swift}.

\bibliographystyle{apj}

\clearpage

\end{document}